\documentclass[eqsecnum,showpacs,preprint,amsmath,aps,nofootinbib]{revtex4}  
\usepackage{epsfig}
\usepackage{amssymb}
\usepackage[dvips]{color}
\usepackage[latin1]{inputenc}
\usepackage{graphicx}
\usepackage{gensymb}
\usepackage{amsmath}
\usepackage{subfig}
\usepackage{relsize}

\def\beq{\begin{equation}}
\def\eeq{\end{equation}}
\def\beqq{\begin{eqnarray}}
\def\eeqq{\end{eqnarray}}

\newcommand{\bdm}{\begin{displaymath}}
\newcommand{\edm}{\end{displaymath}}

\def\pmb#1{\setbox0=\hbox{$#1$}%
  \kern-.025em\copy0\kern-\wd0
  \kern.05em\copy0\kern-\wd0
  \kern-.025em\raise.0433em\box0}

\makeatletter
\renewcommand*{\@fnsymbol}[1]{\ensuremath{\ifcase#1\or *\or \dagger\or
    \ddagger\or 
   \mathsection\or **\or \dagger\dagger
   \or \ddagger\ddagger \else\@ctrerr\fi}}
\makeatother

\begin{document}
\title{Earth-Moon Lagrangian points 
as a testbed for general relativity and effective field theories of gravity}

\author{Emmanuele Battista}
\email[E-mail: ]{ebattista@na.infn.it}
\affiliation{Dipartimento di Fisica, Complesso Universitario 
di Monte S. Angelo, Via Cintia Edificio 6, 80126 Napoli, Italy\\
Istituto Nazionale di Fisica Nucleare, Sezione di Napoli, Complesso Universitario di Monte
S. Angelo, Via Cintia Edificio 6, 80126 Napoli, Italy}

\author{Simone Dell'Agnello}
\email[E-mail: ]{simone.dellagnello@lnf.infn.it}
\affiliation{Istituto Nazionale di Fisica Nucleare, Laboratori Nazionali di Frascati,
00044 Frascati, Italy}

\author{Giampiero Esposito}
\email[E-mail: ]{gesposit@na.infn.it}
\affiliation{Istituto Nazionale di Fisica Nucleare, Sezione di
Napoli, Complesso Universitario di Monte S. Angelo, 
Via Cintia Edificio 6, 80126 Napoli, Italy}

\author{Jules Simo}
\email[E-mail: ]{jules.simo@strath.ac.uk}
\affiliation{Department of Mechanical and Aerospace Engineering, University of Strathclyde, Glasgow, G1 1XJ, United Kingdom\\
School of Engineering, University of Central Lancashire, Preston, PR1 2HE, United Kingdom}

\author{Luciano Di Fiore}
\email[E-mail: ]{luciano.difiore@na.infn.it}
\affiliation{Istituto Nazionale di Fisica Nucleare, Sezione di Napoli,
Complesso Universitario di Monte S. Angelo,
Via Cintia Edificio 6, 80126 Napoli, Italy}

\author{Aniello Grado}
\email[E-mail: ]{aniello.grado@gmail.com}
\affiliation{INAF, Osservatorio Astronomico di Capodimonte, 80131 Napoli, Italy}

\date{\today}

\begin{abstract}
We first analyse the restricted four-body problem consisting of the Earth, the Moon and the 
Sun as the primaries and a spacecraft as the planetoid. This scheme allows us to take into account the solar 
perturbation in the description of the motion of a spacecraft in the vicinity of the stable Earth-Moon 
libration points $L_4$ and $L_5$ both in the classical regime and in the context of effective field theories 
of gravity. A vehicle initially placed at $L_4$ or $L_5$ will not remain near the respective points. 
In particular, in the classical case the vehicle moves on a trajectory about the libration points for at 
least $700$ days before escaping away. We show that this is true also if the modified long-distance Newtonian
potential of effective gravity is employed. We also evaluate 
the impulse required to cancel out the perturbing force due to the Sun in order to force the spacecraft to 
stay precisely at $L_4$ or $L_5$. It turns out that this value is slightly modified with respect to the 
corresponding Newtonian one. In the second part of the paper, we first evaluate the location of all Lagrangian points 
in the Earth-Moon system within the framework of general relativity. For the points $L_4$ and $L_5$, the 
corrections of coordinates are of order a few millimeters and describe a tiny departure from the 
equilateral triangle. After that, we set up a scheme where the theory which is quantum corrected has as its 
classical counterpart the Einstein theory, instead of the Newtonian one. In other words, we deal with a theory 
involving quantum corrections to Einstein gravity, rather than to Newtonian gravity. 
By virtue of the effective-gravity correction to the long-distance form of the potential among two
point masses, all terms involving the ratio between the gravitational radius of the primary and its 
separation from the planetoid get modified. Within this framework, 
for the Lagrangian points of stable equilibrium, we find quantum corrections of order two millimeters, whereas
for Lagrangian points of unstable equilibrium we find quantum corrections below a millimeter. 
Finally, general relativity corrections to Newtonian position of collinear Lagrangian points turn out 
to be below the millimiter, whereas on stable equilibrium points they are of order of a few millimiters. 
\end{abstract}

\pacs{04.60.Ds, 95.10.Ce}

\maketitle

\section{Introduction}

In the space surrounding two bodies that orbit about their mutual mass center there are five points where a third 
body will remain in equilibrium under the gravitational attraction of the other two bodies. These points are called 
Lagrangian points in honour of Joseph Lagrange, who discovered them in 1772 while studying the restricted problem formed 
by the Sun-Jupiter system. The discovery of their physical realization, i.e. the Trojan group of asteroids, 
began only in 1906 thanks to the astronomer Max Wolf with the first-seen member of this group, 588 Achilles, 
which is located near the triangular libration point of the Sun-Jupiter system. Today we know that there are $3898$ 
known Trojans at the triangular Lagrangian point $L_4$ and $2049$ at $L_5$ \cite{MPC}. 
In the sixties, simultaneously 
with the increased interest in space explorations, the question of existence of Lagrangian points with respect to 
other primaries, especially for the Earth-Moon system, arose quite naturally. In fact, if there are stable 
stationary solutions for various primary combinations, then from a practical point of view placing observational 
platforms at these points becomes feasible, especially in a really close and accessible system like 
the Earth-Moon system, which is also the most convenient system from an economic point of view. While the 
Sun-Jupiter system clearly possesses a collection of asteroids at the triangular libration points, the ability of 
the Earth-Moon system to collect debris or dust at the corresponding points and in what is called Kordylewski clouds is 
still in question (see Ref. \cite{Freitas} for further details). The major perturbing effect on the Trojans is represented 
by Saturn, while the stabilizing forces come from the Sun and Jupiter. The major perturbation on the Earth-Moon libration 
clouds is the Sun and the stabilizing effects are derived from the Earth and the Moon. This explains why the existence 
of accumulated material at $L_4$ or $L_5$ in the Earth-Moon system is not so obvious. Bodies at the triangular 
libration points of the system consisting of the Sun and another planet would face the perturbations from Jupiter; 
therefore, it is not surprising that the only currently known material accumulation is confined to 
the Sun-Jupiter system, although some asteroids were found also in the Sun-Earth system around the 
libration point $L_4$, as is shown by recent observations \cite{Connors}. 
As far as the collinear Lagrangian points for the Earth-Moon system are concerned, 
we know that $L_1$ allows comparatively easy access to 
Lunar and Earth orbits with minimal change in velocity and has this as an advantage to position a half-way manned space 
station intended to help transport cargo and personnel to the Moon and backwards, whereas $L_2$ would be a good location for 
a communications satellite covering the Moon's far side and would be an ideal location for a propellant depot as part of 
the proposed depot-based space transportation architecture \cite{Zegler}.

Recently, inspired by the works in Refs. \cite{D94,D94b,D94c,MV95,HL95,ABS97,KK02,D03} on effective field theories 
of gravity, some of us \cite{BE14a,BE14b,BEDS15} have applied this theoretical analysis to the macroscopic bodies 
occurring in celestial mechanics \cite{P1890,P1892,Pars65}, especially in the Earth-Moon system. It has been demonstrated 
that in the quantum regime, when only the interaction potential is modified in the Lagrangian of Newtonian 
gravity, the position of collinear Lagrangian points is governed by four algebraic ninth degree equations, 
which reduce to two algebraic fifth degree equations in the classical regime, while the quantum corrected position of the 
noncollinear libration points is described in terms of a pair of quintic equations, which predict that 
the classical equilateral triangle picture is no longer valid in the quantum scheme. For the Earth-Moon system, 
the prediction about the discrepancy between classical and quantum corrected quantities is of the order of millimeters.
This magnitude is comparable with the instrumental accuracy of point-to-point laser Time-of-Flight (ToF) measurements
in space typical of the modern Satellite/Lunar Laser Ranging (SLR/LLR) techniques 
\cite{Altamimi,Pearlman,Bender,Williams,Shapiro,Daa,Martini,Marcha,Marchb,Dab,Dac,Currie,Dad,Vok}).
The full positioning error budget of the orbits of satellites equipped with laser retroreflectors and reconstructed
by laser ranging depends also on other sources of uncertainty (related to the specific orbit, satellite and
retroreflector array), in addition to the pure point-to-point laser ToF instrumental accuracy (related to the network
of laser ranging ground stations of the ILRS, i.e., the International Laser Ranging Service \cite{Pearlman}. The full
positioning error budget can be larger than millimeters.

This is an interesting potentiality, because we are dealing with predictions 
which might become testable in the Earth-Moon system. 
This is a novel feature in the theory of quantum gravity, 
because all other theories are so far unable to produce testable effects 
\citep{Polchinski98,Rovelli04,Penrose73,Espo11,DeWitt1967c,Kiefer12,Bini13,Bini14,Kamen13,Kamen14,Wilczek14}.  
These predictions become more realistic if we include the perturbations due to the gravitational presence of the Sun, 
in other words we have to face up the restricted problem of four bodies, consisting of the Sun, the Earth and the Moon as 
the three primaries and the fourth body (e.g. a laser-ranged test mass, a spacecraft or by exploiting the solar sail 
technology \cite{Simo08,Simo09a,Simo09b,Simo09c,Simo10a,Simo10b,Simo14,McInnes1999,BEDS15}) which has an infinitesimal mass, to avoid 
affecting the motion of the primaries. As we know, in the restricted three-body problem the motion of the two primaries is 
exactly described by the equations of motion governing the two-body problem. Therefore, we may generalize the problem first 
by solving the dynamical equations describing the motion of the three primaries and then by finding the motion of the planetoid 
in the presumably known gravitational field produced by the primaries. Since no closed-form solution is known for the full 
three-body problem, this generalization to the case of four bodies is rather difficult. A practicable possibility consists 
in assuming the motions of the three primaries and, without attempting to establish the exact solution of the equations 
governing these motions, accept an approximate solution. Such an approximation may be, for instance, that the Earth and the 
Moon move in elliptic orbits around their mass center and that the mass center of the Earth-Moon system, in turn, moves in 
elliptic orbit around the Sun. The plane of the orbit of the mass center of the Earth-Moon system, which is called the plane 
of ecliptic, is inclined relative to the plane containing the orbits of the Earth and the Moon. A simpler approximation 
would consist in neglecting the eccentricity of all orbits, i.e. assuming that the Earth, the Moon and their mass center 
have circular orbits. Under these assumptions, the authors of Ref. \cite{Tapley} did show that, although it is widely 
accepted that, with the introduction of the Sun, the points $L_4$ and $L_5$ of the Earth-Moon system cease to be 
equilibrium points, stable motion may be possible in a region around these noncollinear libration points. The term ``stable'' 
here indicates that the planetoid will remain within a certain region for the period of time during which the motion is studied. 
The work in Ref. \cite{Tapley} demonstrated 
that a spacecraft moves on a trajectory around the stable libration points for 
at least $700$ days before the solar influence causes it to move through wide 
departures from the Lagrangian points. Indeed, from the analysis of the plots it does not appear that,
after $700$ days, a limiting value for the envelope is approached. It would be interesting to recover
this feature directly from the solution of the dynamical equations (if they were known), since at the
present state we believe that the form of the equations involved (see Sec. II) does not allow, by itself,
such a deduction. 

The first purpose of our paper consists in showing that this is true also if we assume the quantum 
corrected potential discovered in Refs. \cite{D94,D03}, and Secs. II and III are devoted to this topic. 
All the considerations made in these Sections represent the natural extension of our previous papers, as we 
continue to describe the three-body problem in the context of effective field theories of gravity by adding all 
features that would contribute to make this subject as close as possible to reality, in order to encourage the launch 
of future space missions that could verify the model we are proposing. On the other hand, one has to consider that general 
relativity is currently the most successful gravitational theory describing the nature of space and time, and well 
confirmed by observations. In fact, it has been brightly confirmed by all the so-called ``classical'' tests, i.e. the 
perihelion shift of Mercury, the deflection of light and the Shapiro time delay, and it has also gone through the systematic 
test offered by the binary pulsar system ``PSR 1913+16'', since the orbit decay of this system is perfectly in accordance 
with the theoretical decay due to the emission of gravitational waves, as predicted by general relativity. Furthermore, 
Lagrangian points have recently attracted renewed interests for relativistic astrophysics 
\cite{Asada09,Yamada10,Yamada,Yamada15}, where the position and the stability of Lagrangian points is described within the 
post-Newtonian regime. For all these reasons, we believe that our model is incomplete without a comparison with the 
Einstein theory. Therefore, Sec. IV studies all Lagrangian points within the framework of general relativity, to 
establish the most accurate classical counterpart of the putative quantum framework that we have set up. By taking 
seriously into account the important role played by the Einstein theory within this scheme, in the last part of this paper 
we describe a {\it new} quantum corrected regime where the underlying classical theory is represented by general relativity, rather 
than Newtonian theory. All the considerations made in Refs. \cite{BE14a,BE14b,BEDS15}, in fact, are characterized by the fact 
that the classical theory for which quantum corrections are computed is the Newtonian theory, instead of Einstein's one. 
But, if general relativity is the most successful classical theory of gravitation, then we have to consider a scheme where 
quantum corrections to general relativity are evaluated. This topic is investigated in Sec. V, where we also show that, among 
all quantum coefficients $\kappa_1$ and $\kappa_2$ in the long-distance corrections to the Newtonian potential 
available in literature \cite{D94,D03,BE14a,BE14b,BEDS15}, the most 
suitable ones to describe the gravitational interactions involving (at least) three bodies are those connected to the 
bound-states potential. Finally, conclusions and open problems are discussed in Sec. VI.

\section{The quantum corrected equations of the restricted four-body problem}

\begin{figure} [htbp] 
\includegraphics[scale=0.7]{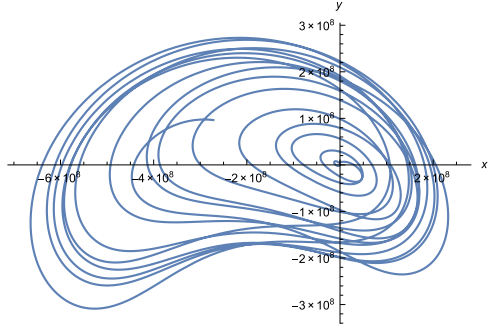}
\caption{Parametric plot of the spacecraft motion about $L_4$ resulting from zero initial displacement and velocity in the 
classical case. The quantities appearing on the axes are measured in meters and the time interval considered is about $4 \times 10^7 {\rm s}$.}
\label{fig3c}
\end{figure}

\begin{figure} [htbp] 
\includegraphics[scale=0.7]{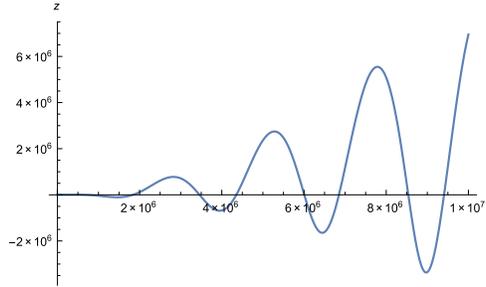}
\caption{Plot of the spacecraft motion about $L_4$ in the $z$-direction resulting from zero initial displacement and 
velocity in the classical case. The quantities on the axes are measured in meters and in seconds.}
\label{fig6c}
\end{figure}

We start by introducing the classical dynamical equations governing the motion of the planetoid in the gravitational field 
of the Earth, the Moon and the Sun \cite{Tapley,Szebehely67}. We suppose that the Earth and the Moon move in circular orbit 
around their mass center and  the mass center, in turn, moves in circular orbit about the Sun. The Earth-Moon orbit plane is 
inclined at an angle $i=5^{\degree} 9^{\prime}$ to the plane of the ecliptic. We introduce the rotating coordinate system 
$\xi,\eta,\zeta $ with the Earth-Moon mass center as its origin and characterized by the fact that the $\xi$ axis lies along 
the Earth-Moon line, the $\eta$ axis lies in the Earth-Moon orbit plane and the $\zeta$ axis points in the direction of the 
angular velocity vector of the Earth-Moon configuration. The $\xi, \eta$ axes rotate about the $\zeta$ axis with the angular 
velocity $\omega$ of the Earth-Moon line. If the vector $\vec{\mathcal{R}}=(\xi,\eta,\zeta)$ indicates in this coordinate 
system the position of a spacecraft of infinitesimal mass, the vector dynamical equation describing its motion is
\begin{equation}
\ddot{\vec{\mathcal{R}}}   + \vec{\omega} \times (2\; \dot{\vec{\mathcal{R}}} + \dot{\vec{\omega}} 
\times \vec{\mathcal{R}}) = -\vec{\nabla}_{\mathcal{R}}V+\vec{\nabla}_{\mathcal{R}}U+ \vec{S}, 
\label{vecc}
\end{equation}
where
\begin{eqnarray}
&V \equiv \dfrac{Gm_1}{\rho_1}+\dfrac{Gm_2}{\rho_2}, \\
&U \equiv Gm_3 \left[\dfrac{1}{\rho_3}-\dfrac{\vec{\mathcal{R}} \cdot 
\vec{\mathcal{R}_3}}{(\mathcal{R}_{3})^{3}} \right],
\end{eqnarray}
with $G$ being the universal gravitation constant, $m_1$,$m_2$ and $m_3$ the mass 
of the Earth, the Moon and the Sun, respectively, $\rho_1$, $\rho_2$ and $\rho_3$ the 
distances from the planetoid of the Earth, the Moon and the Sun, respectively, 
$\mathcal{R}_3$ the distance of the Sun from the Earth-Moon mass center, and lastly $\vec{S}$ 
describes the solar radiation pressure. Written in components, Eq. (\ref{vecc}) becomes
\begin{eqnarray}
\ddot{\xi}-2\omega \dot{\eta}-\omega^2 \xi &= -\dfrac{\partial V}{\partial \xi}
+\dfrac{\partial U}{\partial \xi}+S_{\xi}, 
\label{1ac} \\
\ddot{\eta}+2 \omega \dot{\xi}-\omega^2 \eta &=  -\dfrac{\partial V}{\partial \eta}
+\dfrac{\partial U}{\partial \eta}+S_{\eta},  
\label{1bc} \\
\ddot{\zeta} &=  -\dfrac{\partial V}{\partial \zeta}+\dfrac{\partial U}{\partial \zeta}+S_{\zeta}. 
\label{1cc}
\end{eqnarray}
We can write Eqs (\ref{1ac})--(\ref{1cc}) in what we denote by $x,y,z$ system, which is the 
rotating noninertial coordinate frame of reference centered
at one of the two noncollinear Lagrangian points, e.g. $L_4$. If we use the transformations
\begin{eqnarray}
\xi &=&x+\xi_p, \nonumber \\
\eta &=&y+\eta_p,   \label{3} \\
\zeta &=& z, \nonumber
\end{eqnarray}
where $\xi_p$ and $\eta_p$ are the constant coordinates of the libration point $L_4$ in the 
$\xi,\eta,\zeta$ system, then Eqs. (\ref{1ac})--(\ref{1cc}) become
\begin{equation}
\ddot{x} =2 \omega \dot{y} + (x+\xi_p)\omega^2-(x_3+\xi_p)(\Omega_{\omega})^2 +
S_x + \sum_{i=1}^3 \dfrac{Gm_i}{\rho_{i}^{3}}(x_i-x), 
\label{4ac}
\end{equation}
\begin{equation}
\ddot{y}=-2 \omega \dot{x} + (y+\eta_p)\omega^2-(y_3+\eta_p)(\Omega_{\omega})^2 +
S_y + \sum_{i=1}^3 \dfrac{Gm_i}{\rho_{i}^{3}}(y_i-y), 
\end{equation}
\begin{equation}
\ddot{z}=-z_3(\Omega_{\omega})^2 + S_z + \sum_{i=1}^3 \dfrac{Gm_i}{\rho_{i}^{3}}(z_i-z), \label{4cc}
\end{equation}
where $\Omega_{\omega}$ is the angular velocity of the Earth-Moon mass center around the Sun, and the relation 
$Gm_3/(\mathcal{R}_3)^3=(\Omega_{\omega})^2$ has been exploited. Moreover, the distances $\rho_i$ are given by
\begin{equation}
(\rho_{i})^2=(x_i-x)^2+(y_i-y)^2+(z_i-z)^2 \; \; \; \; \; \; \; (i=1,2,3),
\end{equation}
where the coordinates $(x_1,y_1)$ and $(x_2,y_2)$ of the Earth and the Moon respectively are deduced from 
(\ref{3}) once the coordinates $(\xi_p,\eta_p)$ of $L_4$ are known (remember we have $z_1=z_2=0$), whereas the 
coordinates of the Sun are given by the relations
\begin{eqnarray}
x_3 &=& \mathcal{R}_3 \left( \cos \psi \cos \theta + \cos i \sin \psi \sin \theta \right) -\xi_p, \nonumber \\
y_3 &=& -\mathcal{R}_3 \left( \cos \psi \sin \theta - \cos i \sin \psi \cos \theta \right) -\eta_p, \\
z_3 &=& \mathcal{R}_3 \sin \psi \sin i, \nonumber 
\end{eqnarray}
where $\psi$ is the angular position of the Sun with respect to the vernal equinox 
and measured in the plane of the ecliptic, 
and $\theta$ describes the position of the Earth-Moon line with respect to the vernal equinox measured in the Earth-Moon 
orbit plane (see Fig. $2$ of Ref. \cite{Tapley}). The relations defining these angles are
\begin{gather}
\begin{aligned}
\psi &=& \Omega_{\omega} t + \psi_0,  \\
\theta &=& \Omega_{\omega} t +\theta_0,
\end{aligned}
\end{gather}
where $\psi_0$ and $\theta_0$ are the initial values of $\theta$ and $\psi$. For our computation we have used the following 
numerical values: $\Omega_{\omega}=1.99082 \times 10^{-7}$ rad/s, 
$\omega=2.665075637 \times 10^{-6}$ rad/s, $\psi_0 = \theta_0 = 0$ (i.e. the initial position of the Sun
will be on the extended Earth-Moon line, with the Moon in between Earth and Sun). 
Moreover, following Ref. \cite{BEDS15}, we have the classical values $\xi_p=1.87528148802 \times 10^8$ m and 
$\eta_p = 3.32900165215 \times 10^8 $ m. If we initially set $\vec{S}= \vec{0}$ in (\ref{vecc}), we obtain that the 
perturbative effect of the Sun makes the spacecraft ultimately escape from the stable equilibrium point after about $700$ 
days \cite{Tapley}, as is shown in Figs. \ref{fig3c} and  \ref{fig6c}.
As can be noticed from Fig. \ref{fig3c}, the irregular initial motion damps out and there is an approximate one-month 
periodicity associated with the motion. Moreover, Fig. \ref{fig6c} shows that the amplitude of the motion increases 
with time and that the period of motion is about $27.6$ days, a value really near to the $29.53$ days of the synodical month. 
All these results indicate that the spacecraft will escape from the equilibrium point $L_4$ (or equivalently $L_5$) or, 
in other words, the perturbing presence of the Sun makes the points $L_4$ and $L_5$ cease to be equilibrium points, but they 
are ``stable'' in the sense indicated in the Introduction.

All these considerations are valid within the classical scheme, whereas in the quantum corrected regime 
(see Secs. IV and V) the Newtonian potential is corrected by a Poincar\'e asymptotic expansion involving
integer powers of $G$ only, so that Eq. (\ref{vecc}) can 
be replaced by the vector dynamical equation
\begin{equation}
\ddot{\vec{\mathcal{R}}}   + \vec{\omega} \times (2\; \dot{\vec{\mathcal{R}}} + \dot{\vec{\omega}} 
\times \vec{\mathcal{R}}) = -\vec{\nabla}_{\mathcal{R}}V_{q}+\vec{\nabla}_{\mathcal{R}}U_{q}+ \vec{S}, \label{vecq}
\end{equation}
with \cite{D94,D03,BE14a,BE14b,BEDS15}
\begin{eqnarray}
&V_{q}=\dfrac{Gm_1}{\rho_1} \left[ 1+\dfrac{k_1}{\rho_1}+\dfrac{k_2}{(\rho_1)^2} \right] 
+\dfrac{Gm_2}{\rho_2}\left[ 1+\dfrac{k^{\prime}_{1}}{\rho_2}+\dfrac{k_2}{(\rho_2)^2} \right], \\
&U_{q}=\dfrac{Gm_3}{\rho_3} \left[ 1+\dfrac{k^{\prime \prime}_{1}}{\rho_3}+\dfrac{k_2}{(\rho_3)^2} \right]
-G m_3 \dfrac{\vec{\mathcal{R}} \cdot \vec{\mathcal{R}_3}}{(\mathcal{R}_{3})^{3}} 
\left[1+\dfrac{2 k^{\prime \prime}_{1}}{\mathcal{R}_{3}} +\dfrac{3 k_2}{(\mathcal{R}_{3})^2}\right],
\end{eqnarray}
where, following Ref. \cite{D03}, we decide to adopt the results concerning the bound-states potential. 
Even without knowing the detailed calculations of Sec. V, we may point out that, in classical gravity,
the Levi-Civita cancellation theorem \cite{LeviCivita} holds, according to which the $N$-body Lagrangian
in general relativity can be always reduced to a Lagrangian of $N$ material points. In other words, 
it is not necessary to assume that we deal with point particles for simplicity, but the effects of their
size get eventually and exactly cancelled. Now the quantum corrections considered in Refs.
\cite{BE14a,BE14b,BEDS15} deal precisely with the long-distance Newtonian potential among two such masses,
and consider three distinct physical settings: scattering, or bound states, or one-particle reducible
\cite{D03}. We think that, in celestial mechanics, the bound states picture is more appropriate for
studying stable and unstable equilibrium points. Therefore we set (cf. Sec. V)
\begin{equation}
k_1=-\dfrac{ G m_1}{2 c^2}, \;
k^{\prime}_{1}=-\dfrac{ G m_2}{2 c^2}, \;
k^{\prime \prime}_{1}=-\dfrac{ G m_3}{2 c^2}, \;
k_2=\dfrac{41}{10 \pi} (l_{P})^{2},
\end{equation}
$l_P$ being the Planck length. The occurrence of the term $k_{2}$, which is quadratic in the
Planck length, cannot be obvious for the general reader, and hence we here summarize its properties
and derivation, following our sources \cite{D94,D94b,D94c,D03}. The one-loop quantum correction to the 
gravitational potential is a low-energy property independent of the ultimate high-energy theory. The 
potential of gravitational scattering of two heavy masses turns out to be
$$
V(r)=-{GMm \over r}\left[1+3{G(M+m)\over c^{2}r}+{k_{2}\over r^{2}}\right].
$$
From dimensional analysis one can indeed expect a term like ${k_{2}\over r^{2}}$, because the unique
dimensionless term linear in $\hbar$ and linear in $G$ is ${G \hbar \over c^{3}r^{2}}$. The classical
post-Newtonian correction is also a well-known dimensionless combination, without $\hbar$. We have
a numerical factor of $-{1 \over 2}$ in (2.17) obtained as $3-{7 \over 2}=-{1 \over 2}$, because the
bound-state contribution \cite{D03} written within square brackets above is $-{7 \over 2}$. By
Fourier transform, the corresponding results in momentum space turn out to be \cite{D94c}
$$
{1 \over r} \rightarrow {1 \over q^{2}}, \;
{1 \over r^{2}} \rightarrow {1 \over q^{2}} \times \sqrt{q^{2}}, \;
{1 \over r^{3}} \rightarrow {1 \over q^{2}} \times q^{2} \log(q^{2}), \;
\delta^{3}({\vec x}) \rightarrow {1 \over q^{2}} \times q^{2}.
$$
The one-loop potential is obtained from one-graviton exchange, with the ${1 \over q^{2}}$ 
resulting from the massless propagator. The corrections linear in $\hbar$ are due to all 
one-loop diagrams that can contribute to the scattering of two masses. The kinematic dependence
of the loops then brings in nonanalytic corrections of the form $Gm \sqrt{q^{2}},
Gq^{2}\log(q^{2})$, as well as analytic terms $Gq^{2}$. However, the Fourier transform of the
analytic term is a Dirac delta in position space, and hence analytic terms do not contribute to
long-distance modifications of the potential. The above correspondences are made precise by the
following integrals \cite{D03}:
$$
\int {{\rm d}^{3}q \over (2\pi)^{3}}
{\rm e}^{{\rm i}{\vec q}\cdot {\vec r}}
{1 \over |{\vec q}|^{2}}={1 \over 4 \pi r}, \;
\int {{\rm d}^{3}q \over (2\pi)^{3}}
{\rm e}^{{\rm i}{\vec q}\cdot {\vec r}}
{1 \over |{\vec q}|}={1 \over 2 \pi^{2}r^{2}}, \;
\int {{\rm d}^{3}q \over (2\pi)^{3}}
{\rm e}^{{\rm i}{\vec q}\cdot {\vec r}}
\log(|{\vec q}|^{2})=-{1 \over 2 \pi r^{3}}.
$$

\begin{figure}  
\includegraphics[scale=0.7]{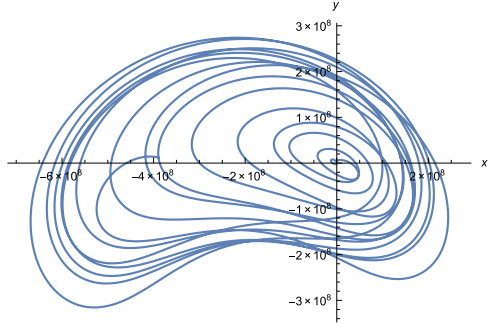}
\caption{Parametric plot of the spacecraft motion about $L_4$ resulting from zero initial displacement and velocity in 
the quantum case. The quantities appearing on the axes are measured in meters and the time interval considered is about $4 \times 10^7 {\rm s}$.}
\label{fig3q}
\end{figure}

\begin{figure}  
\includegraphics[scale=0.7]{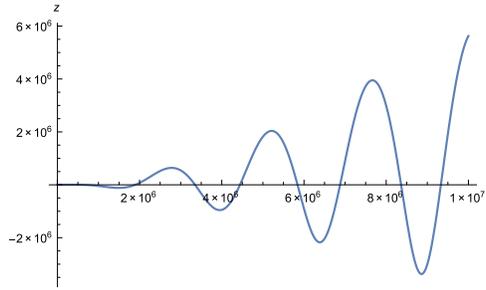}
\caption{Plot of the spacecraft motion about $L_4$ in the $z$-direction resulting from zero initial displacement 
and velocity in the quantum case. The quantities on the axes are measured in meters and in seconds.}
\label{fig6q}
\end{figure}

In the $x,y,z$ system, instead of Eqs. (\ref{4ac})--(\ref{4cc}), 
Eq. (\ref{vecq}), written in components, gives rise to the system
\begin{eqnarray}
\ddot{x} &=& \; 2 \omega \dot{y} + (x+\xi_p)\omega^2-(x_3+\xi_p)(\Omega_{\omega})^2 
\left[1+\dfrac{2 k^{\prime \prime}_{1}}
{\mathcal{R}_{3}} +\dfrac{3 k_2}{(\mathcal{R}_{3})^2}\right]  +\dfrac{G m_1 (x_1-x)}{(\rho_1)^3} 
\left[ 1+\dfrac{2 k_1}{\rho_1}+\dfrac{3 k_2}{(\rho_1)^2} \right] 
\nonumber \\
& + & \dfrac{G m_2 (x_2-x)}{(\rho_2)^3} \left[ 1+\dfrac{2 k^{\prime}_{1}}{\rho_2}+\dfrac{3 k_2}{(\rho_2)^2} \right]  
+ \dfrac{G m_3 (x_3-x)}{(\rho_3)^3} \left[1+\dfrac{2 k^{\prime \prime}_{1}}{\rho_{3}} 
+\dfrac{3 k_2}{(\rho_{3})^2}\right]+ S_x , 
\label{4aq}
\end{eqnarray}
\begin{eqnarray}
\ddot{y} &=& -2 \omega \dot{x} + (y+\eta_p)\omega^2-(y_3+\eta_p)(\Omega_{\omega})^2 
\left[1+\dfrac{2 k^{\prime \prime}_{1}}{\mathcal{R}_{3}} +\dfrac{3 k_2}{(\mathcal{R}_{3})^2}\right]  
+\dfrac{G m_1 (y_1-y)}{(\rho_1)^3} \left[ 1+\dfrac{2 k_1}{\rho_1}+\dfrac{3 k_2}{(\rho_1)^2} \right] 
\nonumber \\
&+& \dfrac{G m_2 (y_2-y)}{(\rho_2)^3} \left[ 1+\dfrac{2 k^{\prime}_{1}}{\rho_2}+\dfrac{3 k_2}{(\rho_2)^2} \right]  
+ \dfrac{G m_3 (y_3-y)}{(\rho_3)^3} \left[1+\dfrac{2 k^{\prime \prime}_{1}}{\rho_{3}} 
+\dfrac{3 k_2}{(\rho_{3})^2}\right]+ S_y , 
\end{eqnarray}
\begin{eqnarray}
\ddot{z} &=& -z_3 (\Omega_{\omega})^2 \left[1+\dfrac{2 k^{\prime \prime}_{1}}{\mathcal{R}_{3}} 
+\dfrac{3 k_2}{(\mathcal{R}_{3})^2}\right]  -\dfrac{G m_1 z}{(\rho_1)^3} \left[ 1+\dfrac{2 k_1}{\rho_1}
+\dfrac{3 k_2}{(\rho_1)^2} \right]  - \dfrac{G m_2 z}{(\rho_2)^3} \left[ 1+\dfrac{2 k^{\prime}_{1}}{\rho_2}
+\dfrac{3 k_2}{(\rho_2)^2} \right] 
\nonumber \\
&+& \dfrac{G m_3 (z_3-z)}{(\rho_3)^3} \left[1+\dfrac{2 k^{\prime \prime}_{1}}{\rho_{3}} 
+\dfrac{3 k_2}{(\rho_{3})^2}\right]+ S_z, 
\label{4cq} 
\end{eqnarray}
where we have used the fact that $z_1=z_2=0$. Setting $\vec{S}=\vec{0}$, we have integrated Eqs. (\ref{4aq})--(\ref{4cq}) 
and we have discovered that the situation is almost the same as in the classical case (see Figs. \ref{fig3q} and \ref{fig6q}), 
i.e. the planetoid is destined to run away from the triangular libration points in about $700$ days. This means that, 
also within a quantum corrected scheme, the gravitational effect of the Sun spoils the equilibrium condition at $L_4$ and $L_5$.  

\section{The solar radiation pressure and the linear stability at $L_4$}

\begin{figure}  
\includegraphics[scale=0.7]{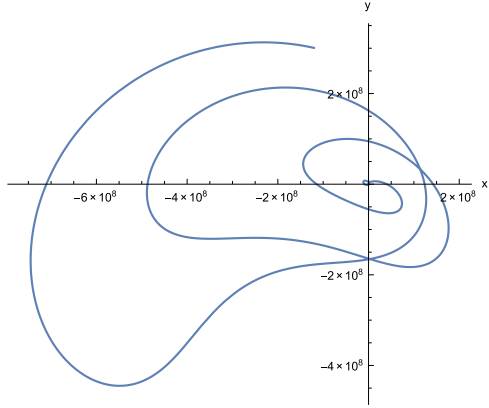}
\caption{Parametric plot in the classical regime of the spacecraft motion about $L_4$ in the presence of the solar radiation 
pressure and considering $A/m=0,159\; {\rm m^2/Kg}$. The initial displacement and velocity are zero. The quantities appearing 
on the axes are measured in meters and the time interval considered is about $1 \times 10^7 {\rm s}$.}
\label{fig8c}
\end{figure}

At this stage, we assume the presence of the radiation pressure both in the classical equations (\ref{4ac})--(\ref{4cc}) 
and in the quantum ones (\ref{4aq})--(\ref{4cq}). The solar radiation pressure is given by
\begin{equation}
\vec{S}= -K \dfrac{A}{m (\rho_3)^3} \vec{\rho}_3, \label{solar_radiation}
\end{equation}  
where $A$ is the cross-sectional area normal to $\vec{\rho}_3$, $m$ is the planetoid mass and $K$ is a constant. 
Inspired by Ref. \cite{Tapley}, we use the value $K=2,048936 \times 10^{17} N$. We have integrated the 
classical equations (\ref{4ac})--(\ref{4cc}) and we have found that the presence of the solar 
radiation pressure causes the vehicle to move further away from $L_4$ in a given time, as one can see from Fig. \ref{fig8c}. 
In particular, the larger the ratio $A/m$ is, the larger the envelope of the motion is \cite{Tapley}. 

\begin{figure}  
\includegraphics[scale=0.6]{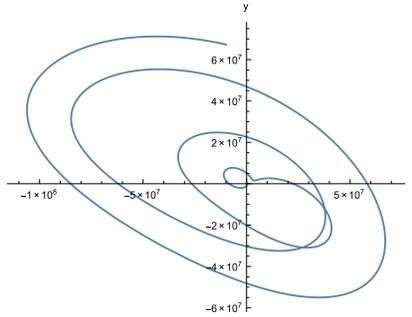}
\caption{Parametric plot in the quantum regime of the spacecraft motion about $L_4$ in the presence of the solar radiation 
pressure and considering $A/m=0,159\; {\rm m^2/Kg}$. The initial displacement and velocity are zero. The quantities appearing 
on the axes are measured in meters and the time interval considered is about $1 \times 10^7 {\rm s}$.}
\label{fig8q}
\end{figure}

\begin{figure}
\[
\begin{array}{cccc}
\includegraphics[scale=0.7]{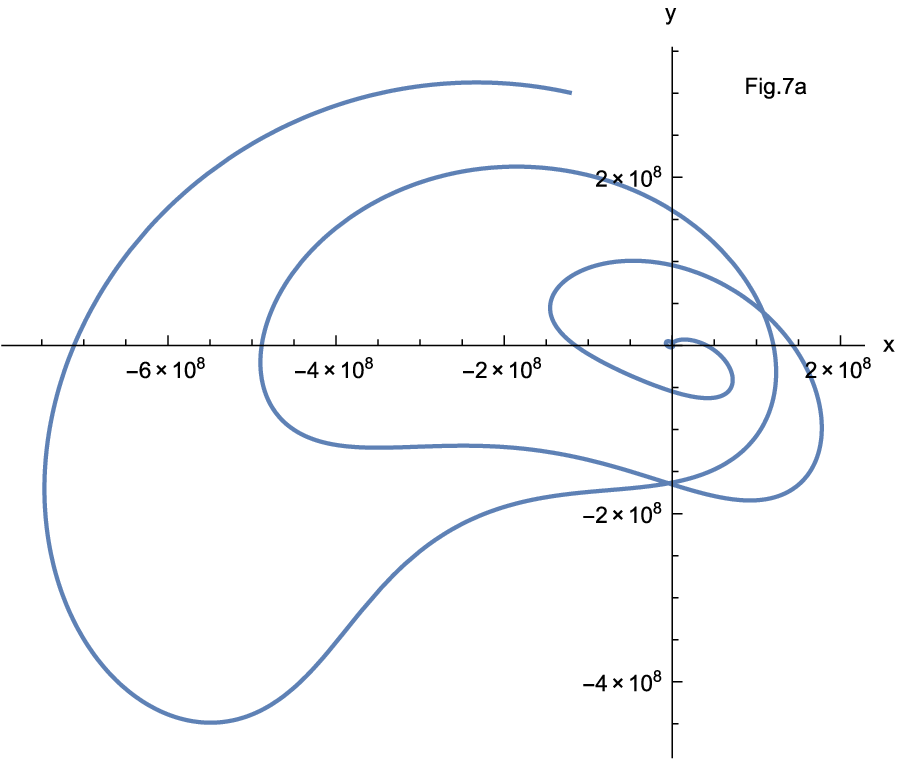}  
& \includegraphics[scale=0.7]{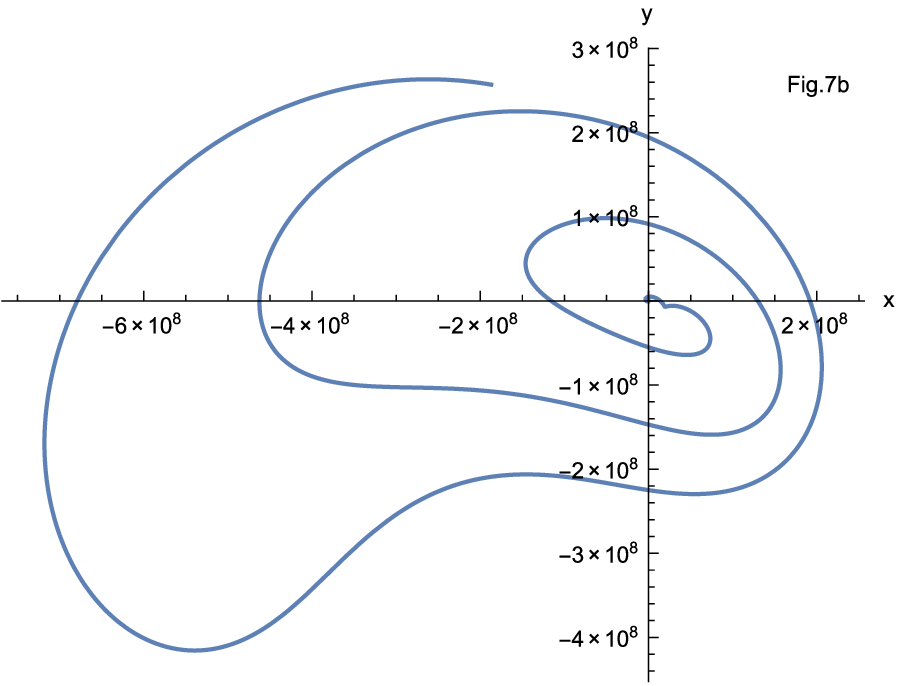} \\
\includegraphics[scale=0.7]{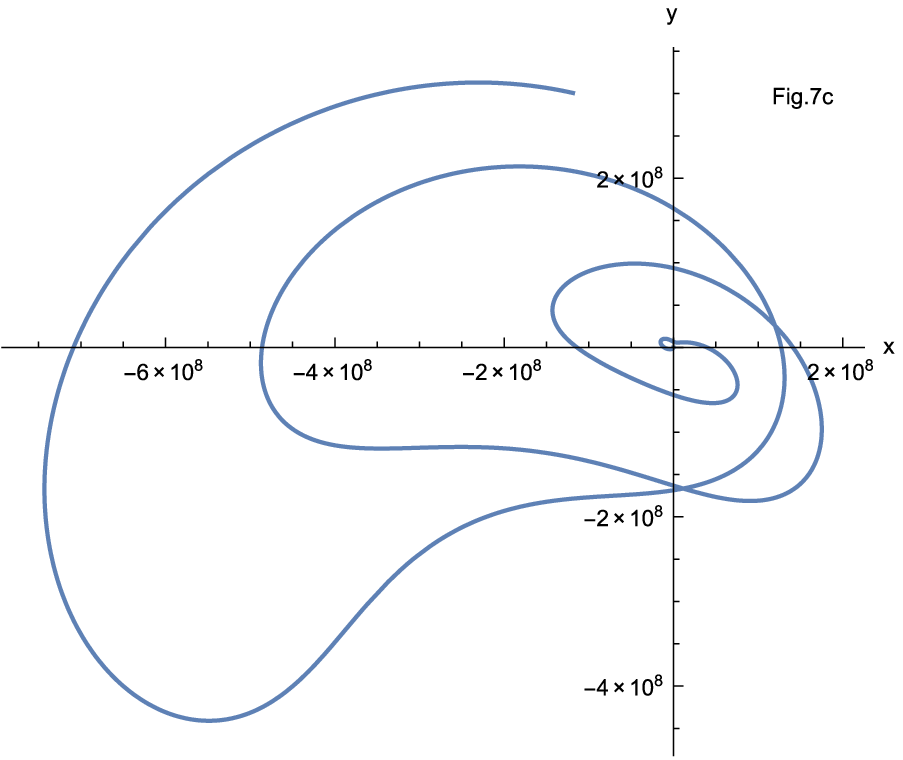}
& \includegraphics[scale=0.7]{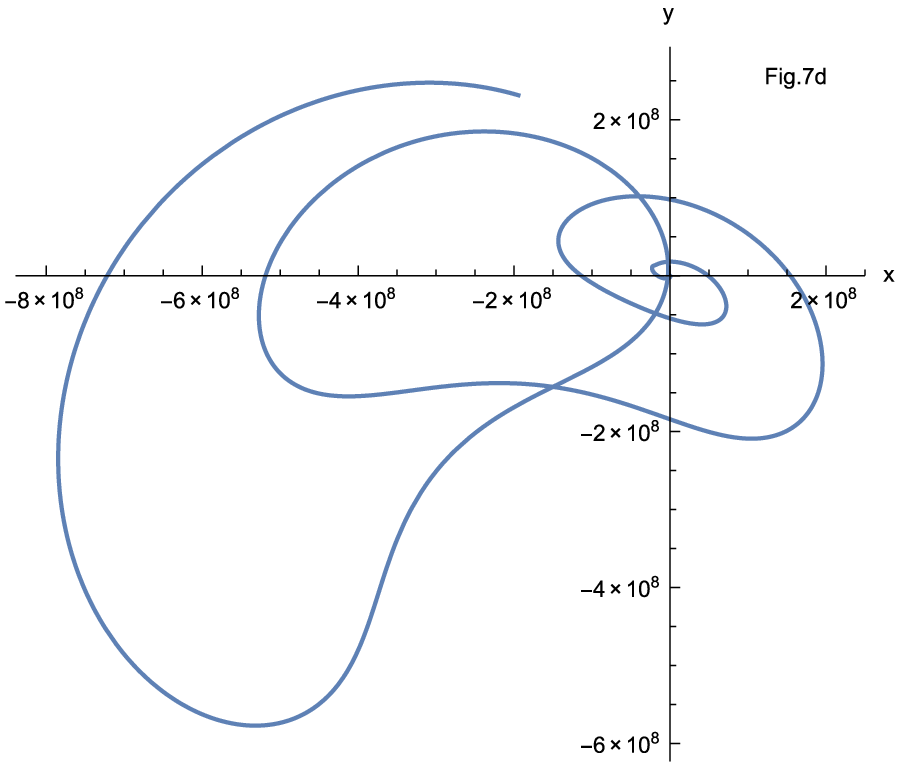}
\cr
\end{array}
\]
\caption{Fig. 7a: Spacecraft motion about $L_4$ in the classical regime and with an initial velocity 
of $3\; {\rm m/s}$ at $60^{\degree}$; Fig. 7b: Spacecraft motion about $L_4$ in the classical regime and 
with an initial velocity of $3\; {\rm m/s}$ at $150^{\degree}$; Fig. 7c:
Spacecraft motion about $L_4$ in the classical regime and with an initial velocity of $3\; {\rm m/s}$ at $240^{\degree}$;
Fig. 7d: Spacecraft motion about $L_4$ in the classical regime and with an initial velocity 
of $3\; {\rm m/s}$ at $330^{\degree}$.}
\label{fig12c}
\end{figure}

Interestingly, in the quantum case ruled by effective gravity 
the situation is a little bit different. Unlike the classical regime, the presence of the 
solar radiation pressure in the equations (\ref{4aq})--(\ref{4cq}) does not show itself through the fact that the spacecraft 
goes away from the triangular libration points more rapidly, but it results in a less chaotic and irregular motion about 
$L_4$, which ultimately make the planetoid escape from $L_4$, like in the classical case. These effects are clearly visible from Fig. 
\ref{fig8q}\footnote{The different scale adopted in Fig. \ref{fig8q} with respect to the one of Fig. \ref{fig8c} allows us to better 
appreciate its features.}.
We can also try to find the best set of initial conditions which leads to the smallest envelope of the motion of the 
planetoid. We have studied several sets of initial conditions both in the classical case and in the quantum one. In the 
classical regime, we completely agree with the results of Ref. \cite{Tapley}. We have found in fact that the amplitude of the 
spacecraft's motion depends strongly on the position of the Sun (i.e. on the values assumed by $\theta_0$ and $\psi_0$) and 
on its initial position and velocity. For example, Fig. \ref{fig12c} shows the motion resulting from an initial 
zero displacement and different initial velocity ($\theta_0=\psi_0=0$) and the time interval considered is about $1 \times 10^7 {\rm s}$.
As we can see, the envelope of the motion in Fig. 7b is smaller at any time than the envelope 
of the motion shown in Fig. \ref{fig8c}.

\begin{figure}
\[
\begin{array}{cccc}
\includegraphics[scale=0.7]{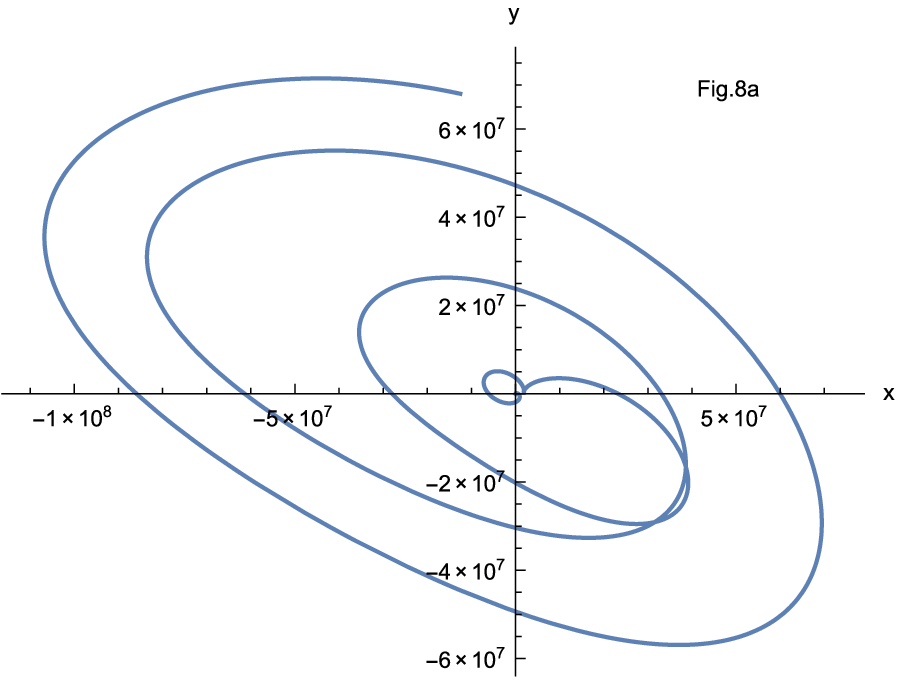} 
& \includegraphics[scale=0.7]{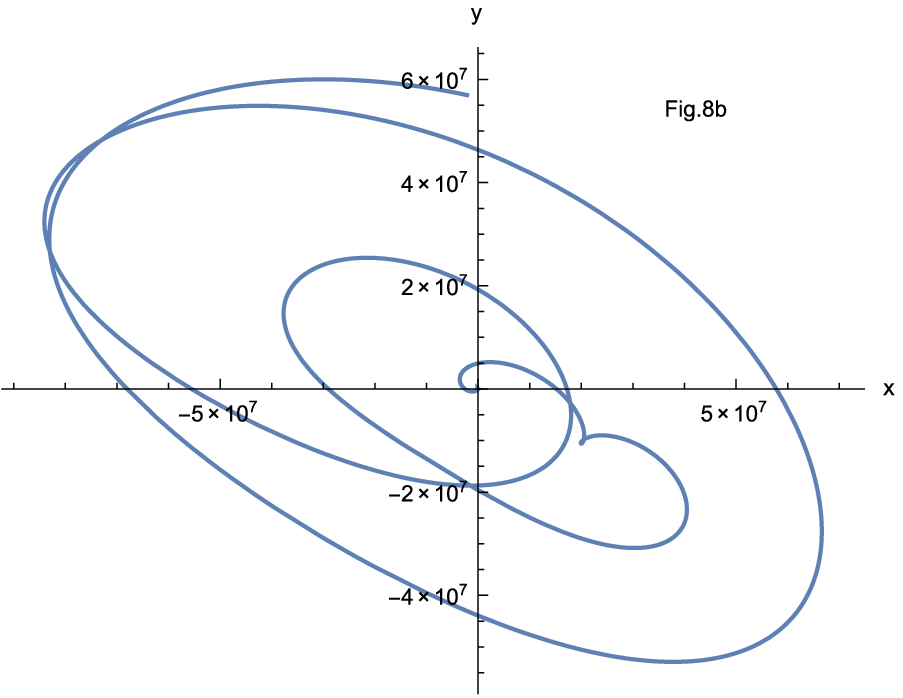} \\
 \includegraphics[scale=0.7]{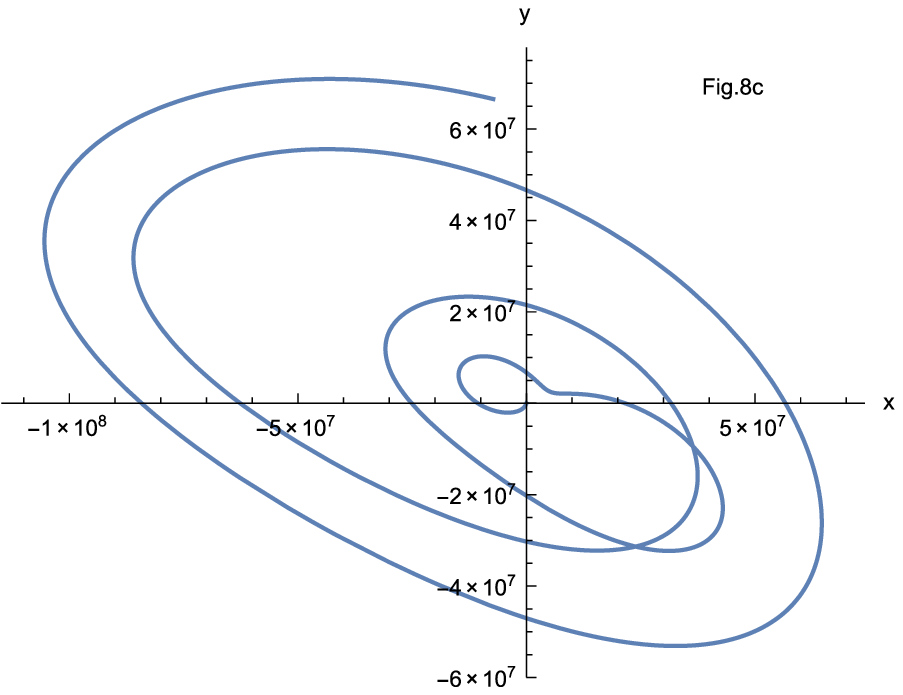} 
& \includegraphics[scale=0.7]{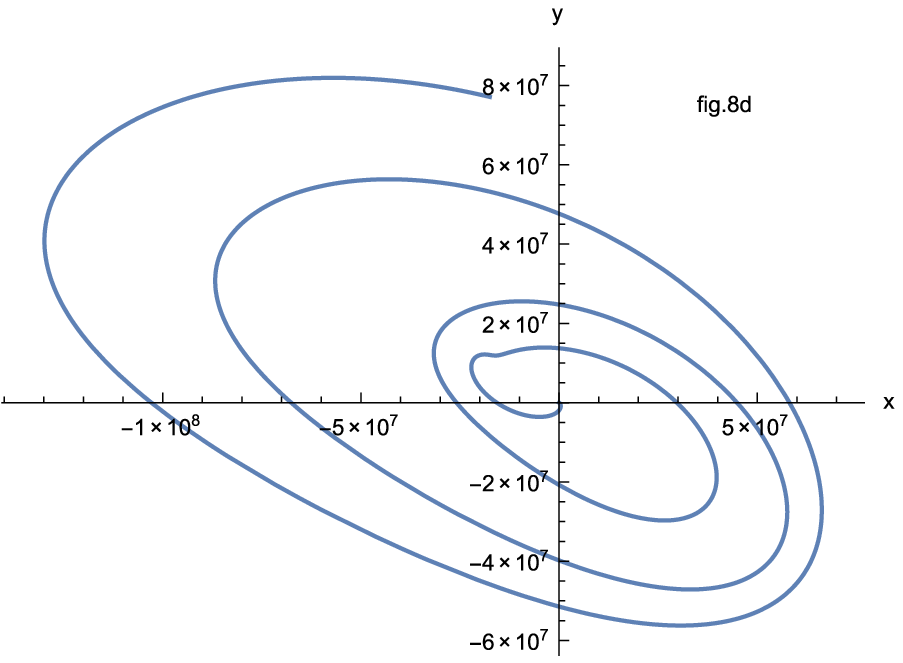}
\cr
\end{array}
\]
\caption{Fig. 8a: Spacecraft motion about $L_4$ in the quantum regime and with an initial velocity of $3\; {\rm m/s}$ at $60^{\degree}$;
Fig. 8b: Spacecraft motion about $L_4$ in the quantum regime and with an initial velocity of $3\; {\rm m/s}$ at $150^{\degree}$;
Fig. 8c: Spacecraft motion about $L_4$ in the quantum regime and with an initial velocity of $3\; {\rm m/s}$ at $240^{\degree}$;
Fig. 8d: Spacecraft motion about $L_4$ in the quantum regime and with an initial velocity of $3\; {\rm m/s}$ at $330^{\degree}$.}
\label{fig12q}
\end{figure}

The situation is fairly the same in the quantum regime (Fig. \ref{fig12q}), where we have discovered that one set of 
initial conditions (having $\theta_0=\psi_0=0$) exists which results in a smaller envelope of the 
spacecraft motion at any given time, as one can see from Fig. 8b. This fact can be understood with a comparison between Figs. 6 
and 8b. The interesting difference with respect to the classical case consists in the fact that 
the reduction of the envelope of the planetoid motion produced by a nonzero initial velocity becomes more evident 
in the quantum regime. By inspection of Figs. 7 and 8 we discover a strong dependence on the initial conditions
of the planetoid trajectories both in the classical and quantum regime. This suggests that, from an experimental
point of view, it might be useful to drop off two or more satellites close to the Lagrangian points $L_{4}$
and $L_{5}$ with slightly different initial conditions for position and velocity. Measurements of the satellite
differential positions, together with the measurement of the single orbits, could make it possible to discriminate
between classical and quantum regime, without depending on the absolute knowledge of Lagrangian points' location.  

\begin{figure}
\[
\begin{array}{cc}
\includegraphics[scale=0.7]{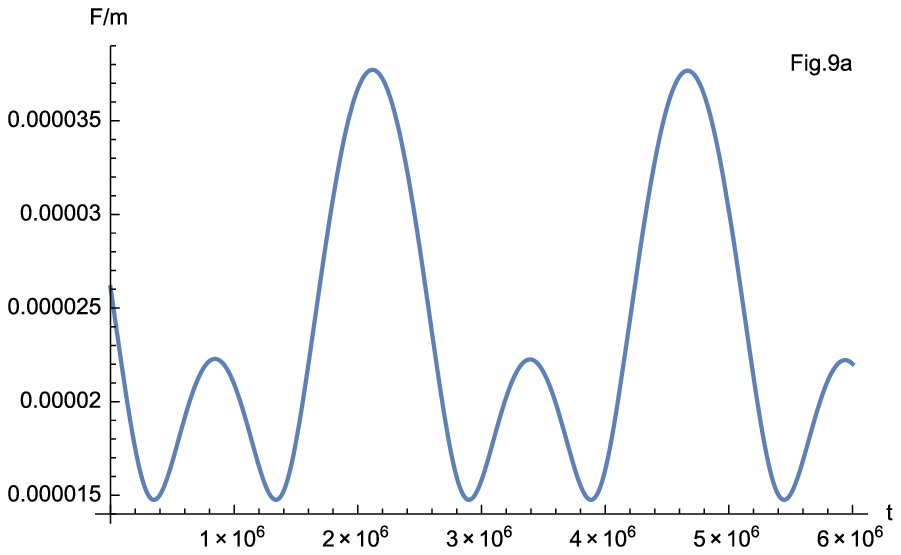} & \includegraphics[scale=0.7]{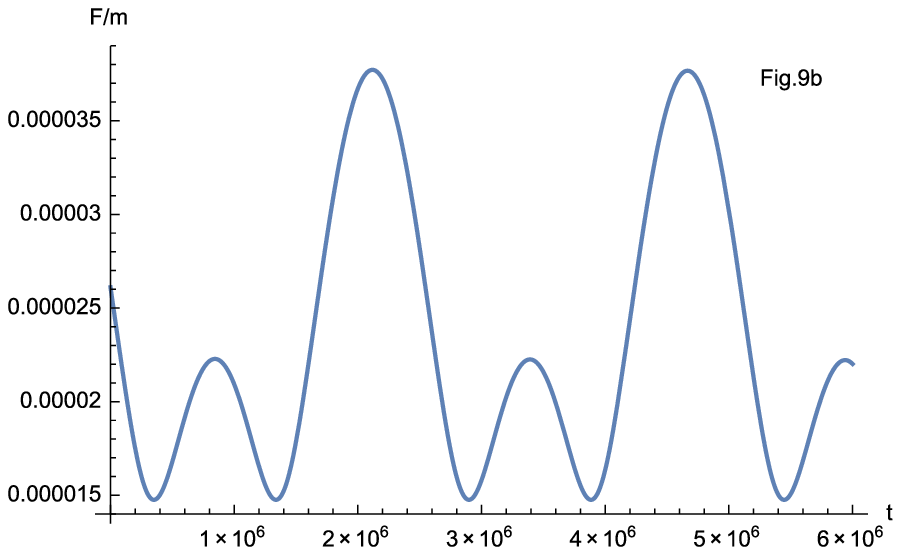} 
\cr
\end{array}
\]
\caption{Fig. 9a: Force per unit mass as a function of time required to induce stability at $L_4$ in the classical regime;
Fig. 9b: Force per unit mass as a function of time required to induce stability at $L_4$ in the quantum regime.}
\label{impulse}
\end{figure}

If we want to force the particle to stay precisely at $L_4$, we have to set aside the perturbing force due 
to the Sun by the application of a continuous force (see Fig. \ref{impulse}).
Therefore, we have to study the following stability equation (in the $\xi,\eta,\zeta$ system):
\begin{equation}
-\vec{\nabla}_{\mathcal{R}}V+\vec{\nabla}_{\mathcal{R}}U+ \vec{S}+\dfrac{\vec{F}}{m}=\vec{0}, \label{vecstabilityc}
\end{equation}
which becomes in the quantum case
\begin{equation}
-\vec{\nabla}_{\mathcal{R}}V_q+\vec{\nabla}_{\mathcal{R}}U_q+ \vec{S}+\dfrac{\vec{F_q}}{m}=\vec{0}, \label{vecstabilityq}
\end{equation}
where $m$ is the mass of the planetoid and $\vec{F}$ 
(respectively, $\vec{F}_q$) represents the force to be 
applied to the spacecraft in order to make it stay precisely at $L_4$ in the 
classical (respectively, quantum) regime. 
If we consider Eqs. (\ref{vecstabilityc})--(\ref{vecstabilityq}) in the $x,y,z$ coordinate system, we can exploit 
the simplification resulting from the fact that the planetoid must be at the position $x=y=z=0$, hence Eq. 
(\ref{vecstabilityc}), written in components, becomes
\begin{equation}
\xi_p \omega^2-(x_3+\xi_p)(\Omega_{\omega})^2 + S_x 
+ \sum_{i=1}^3 \dfrac{Gm_i}{\rho_{i}^{3}} x_i+ \dfrac{F_x}{m}=0, \label{stabilityac}
\end{equation}
\begin{equation}
\eta_p \omega^2-(y_3+\eta_p)(\Omega_{\omega})^2 + S_y 
+ \sum_{i=1}^3 \dfrac{Gm_i}{\rho_{i}^{3}} y_i + \dfrac{F_y}{m}=0, 
\end{equation}
\begin{equation}
-z_{3}(\Omega_{\omega})^2 + S_z + \dfrac{Gm_3}{\rho_{i}^{3}}z_3 + \dfrac{F_z}{m}=0, \label{stabilitycc}
\end{equation}
whereas from Eq. (\ref{vecstabilityq}) we obtain 
\begin{eqnarray}
\; & \; & \xi_p \omega^2-(x_3+\xi_p)(\Omega_{\omega})^2 
\left[1+\dfrac{2 k^{\prime \prime}_{1}}{\mathcal{R}_{3}} 
+\dfrac{3 k_2}{(\mathcal{R}_{3})^2}\right]  +\dfrac{G m_1 x_1}{(\rho_1)^3} 
\left[ 1+\dfrac{2 k_1}{\rho_1}+\dfrac{3 k_2}{(\rho_1)^2} \right] 
\nonumber \\
&+& \dfrac{G m_2 x_2}{(\rho_2)^3} \left[ 1+\dfrac{2 k^{\prime}_{1}}{\rho_2}+\dfrac{3 k_2}{(\rho_2)^2} \right]  
+ \dfrac{G m_3 x_3}{(\rho_3)^3} \left[1+\dfrac{2 k^{\prime \prime}_{1}}{\rho_{3}} +\dfrac{3 k_2}{(\rho_{3})^2}\right]
+ S_x +  \dfrac{F_{q_x}}{m}=0, 
\label{stabilityaq} 
\end{eqnarray}
\begin{eqnarray}
\; & \; &
\eta_p\omega^2-(y_3+\eta_p)(\Omega_{\omega})^2 \left[1+\dfrac{2 k^{\prime \prime}_{1}}{\mathcal{R}_{3}} 
+\dfrac{3 k_2}{(\mathcal{R}_{3})^2}\right]  +\dfrac{G m_{1} y_{1}}{(\rho_1)^3} 
\left[ 1+\dfrac{2 k_1}{\rho_1}+\dfrac{3 k_2}{(\rho_1)^2} \right] 
\nonumber \\
&+& \dfrac{G m_2 y_2}{(\rho_2)^3} \left[ 1+\dfrac{2 k^{\prime}_{1}}{\rho_2}+\dfrac{3 k_2}{(\rho_2)^2} \right]  
+ \dfrac{G m_3 y_3}{(\rho_3)^3} \left[1+\dfrac{2 k^{\prime \prime}_{1}}{\rho_{3}} 
+\dfrac{3 k_2}{(\rho_{3})^2}\right]+ S_y+\dfrac{F_{q_y}}{m}=0, 
\end{eqnarray}
\begin{equation}
-z_{3} (\Omega_{\omega})^2 \left[1+\dfrac{2 k^{\prime \prime}_{1}}{\mathcal{R}_{3}} 
+\dfrac{3 k_2}{(\mathcal{R}_{3})^2}\right]  
+  \dfrac{G m_3 z_3}{(\rho_3)^3} \left[1+\dfrac{2 k^{\prime \prime}_{1}}{\rho_{3}} 
+\dfrac{3 k_2}{(\rho_{3})^2}\right]+ S_z + \dfrac{F_{q_z}}{m}=0. 
\label{stabilitycq}
\end{equation}
Equations (\ref{stabilityac})--(\ref{stabilitycc}) and (\ref{stabilityaq})--(\ref{stabilitycq}) 
make it possible for us to evaluate 
both the classical and the quantum force needed for stability and therefore the impulse per 
unit mass which the planetoid must be subjected to in order to stay in equilibrium exactly at $L_4$. 
Bearing in mind that  the impulse is defined as the integral of a force over the time interval for which it acts, 
and on considering a time interval of one year, we have found the following results for the classical and the 
quantum regime, respectively:
\begin{gather}
\begin{aligned}
I_c /m = 747,608255 \; {\rm N \;s /Kg},  \\
I_q /m = 747,608245 \; {\rm N \; s /Kg}. 
\label{impulsevalue}
\end{aligned}
\end{gather}
Of course, these considerations are preliminary because, even just at classical level, the 
four-body problem has only been studied from us within Newtonian gravity.
We also note that this calculation suggests a gedanken experiment in which
two satellites are sent to $L_{4}$ and $L_{5}$, respectively.
If the first satellite receives the impulse $I_{c}$ while the second receives the impulse $I_{q}$,
one might try to check, by direct comparison, which value is better suited for stabilizing the
Lagrangian point, gaining support for classical or, instead, quantum theory.
However, this configuration is merely ideal because, in light of the very small relative difference
of the impulse in the two cases, it looks practically impossible to keep all the experimental
conditions (satellite mass, actuator and readout calibration, initial conditions, solar
radiation pressure etc.) identical within the required accuracy (less than 0.1 parts
per million). 

\section{Theoretical predictions of general relativity}
\subsection{Noncollinear Lagrangian points}

The analysis of the previous section relies on the simple but nontrivial {\it assumption} that, since
effective gravity modifies the long-distance Newtonian potential among bodies of masses $m_{A}$ and
$m_{B}$ according to\footnote{We can say that the real number $\kappa_{1}$ is the effective-gravity
weight of the sum of gravitational radii, whereas the real number $\kappa_{2}$ is the effective-gravity
weight of Planck's length squared.}
\begin{eqnarray}
\; & \; & 
V_{E}(r) \sim -{G m_{A}m_{B}\over r}\left[ 1+ \left(\kappa_{1}{(R_{A}+R_{B})\over r}
+\kappa_{2}{(l_{P})^{2}\over r^{2}} +{\rm O}(G^{2})\right)\right]
\nonumber \\
& \Longrightarrow & {V_{E}(r) \over c^{2}m_{B}} \sim 
-{R_{A}\over r}\left[ 1+ \left(\kappa_{1}{(R_{A}+R_{B})\over r}
+\kappa_{2}{(l_{P})^{2}\over r^{2}} +{\rm O}(G^{2})\right)\right],
\label{(4.1)}
\end{eqnarray}
for all values of $r$ greater than a suitably large $r_{0}$, where gravitational radii 
$R_{A},R_{B}$ and Planck length $l_{P}$ are defined by
\begin{equation}
R_{A} \equiv {G m_{A}\over c^{2}}, \; R_{B} \equiv {G m_{B} \over c^{2}}, \;
l_{P} \equiv \sqrt{G \hbar \over c^{3}}, 
\label{(4.2)}
\end{equation}
the resulting modification of Newtonian dynamics can be obtained by considering a classical 
Lagrangian where the Newtonian potential 
\begin{equation}
V_{N}(r)=-{G m_{A}m_{B}\over r}
\label{(4.3)}
\end{equation}
is replaced by $V_{E}(r)$, while all other terms remain unaffected (cf. Sec. V). Although
it would be inappropriate to use the quantum effective action to study the low-energy effects resulting
from the asymptotic expansion (\ref{(4.1)}), the above assumption is a shortcut to describe
a theory lying in between classical gravity and full quantum gravity. For this reason, it becomes
important to study the predictions of classical gravity when general relativity is instead assumed.
The work in Ref. \cite{Yamada} has indeed done so by relying upon the Einstein-Infeld-Hoffmann equations
of motion for a three-body system, but without studying an effective potential and the zeros of its 
gradient. However, such a potential is by now available in the literature, and the resulting approximate
evaluation of Lagrangian points $L_{4}$ and $L_{5}$ was performed in Ref. \cite{Bhat}, while their stability
in a suitable mass range was proved in Ref. \cite{Duscos}. Strictly speaking, in general relativity the
libration points become quasi-libration points, and we refer the reader to Ref. \cite{Brumberg} for this
feature, which reflects the expected emission of gravitational radiation.

For our purposes, it is enough to consider the relativistic version of the circular restricted
three-body problem in a plane, where, for primaries 
of masses $\alpha$ and $\beta$ separated by a distance $l$, with
gravitational radii $R_{\alpha} \equiv {G \alpha \over c^{2}}$ and $R_{\beta} \equiv {G \beta \over c^{2}}$,
and mass ratio $\rho \equiv {\beta \over \alpha} <1$, the classical angular frequency (or pulsation)
$\omega \equiv \sqrt{G(\alpha+\beta)\over l^{3}}$ is replaced by \cite{Bhat}
\begin{equation}
\Omega \equiv \omega \left[1-{3 \over 2}{(R_{\alpha}+R_{\beta})\over l}
\left(1-{1 \over 3}{\rho \over (1+\rho)^{2}}\right)\right],
\label{(4.4)}
\end{equation}
while, in a noninertial frame with origin at the mass center of the Earth-Moon system, the equations
of motion of the planetoid in the planar case (with coordinates $(\xi,\eta)$) read as \cite{Bhat}
\begin{equation}
\ddot \xi -2 \Omega {\dot \eta}={\partial W \over \partial \xi}
-{{\rm d}\over {\rm d}t}\left( {\partial W \over \partial {\dot \xi}}\right),
\label{(4.5)}
\end{equation}
\begin{equation}
\ddot \eta +2 \Omega {\dot \xi}={\partial W \over \partial \eta}
-{{\rm d}\over {\rm d}t}\left( {\partial W \over \partial {\dot \eta}}\right),
\label{(4.6)}
\end{equation}
where, upon denoting by $r$ the distance of the planetoid from the primary of mass $\alpha$ (i.e.
the Earth), and by $s$ the distance of the planetoid from the primary of mass $\beta$ (i.e. the Moon),
given by
\begin{equation}
r^{2}=\left(\xi + {\rho l \over (1+\rho)}\right)^{2}+\eta^{2},
\label{(4.7)}
\end{equation}
\begin{equation}
s^{2}=\left(\xi - {l \over (1+\rho)}\right)^{2}+\eta^{2},
\label{(4.8)}
\end{equation}
one has the effective potential reading as \cite{Bhat}
\begin{eqnarray}
W &=& {\Omega^{2}\over 2}(\xi^{2}+\eta^{2})+c^{2}\left[{R_{\alpha}\over r}
+{R_{\beta}\over s}-{1 \over 2}\left({(R_{\alpha})^{2}\over r^{2}}
+{(R_{\beta})^{2}\over s^{2}}\right)\right] 
\nonumber \\
&+& {1 \over 8c^{2}}f^{2}(\xi,\eta,{\dot \xi},{\dot \eta})
+{3 \over 2}\left({R_{\alpha}\over r}+{R_{\beta}\over s}\right)
f(\xi,\eta,{\dot \xi},{\dot \eta})
\nonumber \\
&+& {R_{\beta}\over (1+\rho)}\Omega l \left(4 {\dot \eta}
+{7 \over 2}\Omega \xi \right)\left({1 \over r}-{1 \over s}\right)
\nonumber \\
&+& {R_{\beta}\over (1+\rho)}\Omega^{2}l^{2} \left[-{\eta^{2}\over 2(1+\rho)}
\left({\rho \over r^{3}}+{1 \over s^{3}}\right)-{l \over rs}
+{(\rho-2)\over 2(1+\rho)}{1 \over r}
+{(1-2 \rho)\over 2(1+\rho)}{1 \over s}\right],
\label{(4.9)}
\end{eqnarray}
where \cite{Bhat}
\begin{equation}
f(\xi,\eta,{\dot \xi},{\dot \eta}) \equiv {\dot \xi}^{2}+{\dot \eta}^{2}
+2 \Omega (\xi {\dot \eta}-\eta {\dot \xi})
+\Omega^{2}(\xi^{2}+\eta^{2}).
\label{(4.10)}
\end{equation}
At all equilibrium points, the first and second time derivatives of coordinates $(\xi,\eta)$ should
vanish, which implies that it is enough to evaluate the zeros of the gradient of $W(\xi,\eta)$, 
because \cite{Bhat}
\begin{equation}
{{\rm d}\over {\rm d}t}\left({\partial W \over \partial {\dot \xi}}\right)
={{\rm d}\over {\rm d}t}\left({\partial W \over \partial {\dot \eta}}\right)=0 \; 
{\rm if} \; {\dot \xi}={\dot \eta}={\ddot \xi}={\ddot \eta}=0.
\label{(4.11)}
\end{equation}
Note now that, by virtue of (\ref{(4.7)}) and (\ref{(4.8)}), one has the formulas 
(\ref{(A1)})-(\ref{(A4)}) in the Appendix, and hence
the two components of the gradient can be expressed in the form
\begin{equation}
{\partial W \over \partial \xi}=W_{1}(\xi,\eta,r)+W_{2}(\xi,\eta,s)
+{R_{\beta}l^{3}\over (1+\rho)}{\Omega^{2}\over rs}
\left[\xi \left({1 \over r^{2}}+{1 \over s^{2}}\right)
+{l \over (1+\rho)}\left({\rho \over r^{2}}-{1 \over s^{2}}\right)\right],
\label{(4.12)}
\end{equation}
\begin{equation}
{\partial W \over \partial \eta}=\eta \left[W_{3}(\xi,\eta,r)+W_{4}(\xi,\eta,s)
+{R_{\beta}l^{3}\over (1+\rho)}{\Omega^{2}\over rs}
\left({1 \over r^{2}}+{1 \over s^{2}}\right)\right],
\label{(4.13)}
\end{equation}
where the functions $W_{1},...,W_{4}$ are defined in Eqs. (\ref{(A5)})-(\ref{(A8)}). Thus, unlike the case
of Refs. \cite{BE14a,BE14b,BEDS15}, 
when the gradient of $w$ is set to zero with $\eta \not=0$, one does not
get an algebraic equation for $r$ only. Since we are interested in numerical solutions of such 
an enlarged algebraic system with (at least) ten decimal digits, we set $r \equiv \gamma l$,
$s=\Gamma l$, and we study the coupled algebraic equations for the real numbers $\gamma$ and
$\Gamma$ obtained from
\begin{equation}
\gamma^{5}\Gamma^{5}{\partial W \over \partial \xi}=0,
\label{(4.14)}
\end{equation}
\begin{equation}
\gamma^{5}\Gamma^{5}{1 \over \eta}{\partial W \over \partial \eta}=0,
\label{(4.15)}
\end{equation}
where the fifth powers of $\gamma$ and $\Gamma$ are suggested by the occurrence of terms proportional
to $\gamma^{-5}$ and $\Gamma^{-5}$ in the derivatives ${\partial W \over \partial \xi}$ and
${\partial W \over \partial \eta}$. We can write Eqs.
(\ref{(4.14)})--(\ref{(4.15)}) in a more concise way, i.e.
\begin{equation}
\gamma^{5}\Gamma^{5}{\partial W \over \partial \xi} =\sum_{n=0}^{5}A_n(\Gamma^{j})\gamma^n=0, 
\; \; \; j \in \{0,1,2,3,4,5 \}, 
\label{(4.16)}
\end{equation}
\begin{equation}
\gamma^{5}\Gamma^{5}{1 \over \eta}{\partial W \over \partial \eta} = \sum_{n=0}^{5}B_n(\Gamma^{j})\gamma^n=0, 
\; \; \; j \in \{0,1,2,3,4,5 \}, 
\label{(4.17)}
\end{equation}
where the coefficients $A_n(\Gamma^j)$ are given by
\begin{eqnarray}
A_5(\Gamma^j) & \equiv & \Gamma^5 \left[ 1+\dfrac{\Omega^2}{2c^2} \left(\eta^2+ \xi^2 \right) \right] \xi \Omega^2 
+ \Gamma^4 \left[ \dfrac{3 \xi}{l}-\dfrac{7}{2 \left(1+\rho \right)}\right]R_{\beta} \Omega^2
\nonumber \\
& + & \Gamma^2 \left\{ \frac{1}{2} \Omega^2 \left[ \dfrac{7 l \xi}{\left(1+ \rho \right)} 
+ \dfrac{l^2 \left(2 \rho-1 \right)}{\left(1+ \rho \right)^2} -3 \left(\eta^2 + \xi^2 \right) \right] 
- c^2  \right\} \left(\xi - \dfrac{l}{(1+ \rho)} \right) \dfrac{R_{\beta}}{l^3} 
\nonumber \\
& + & \left[  \Gamma   \left(\dfrac{c R_{\beta} }{l^2} \right)^2  
+ \dfrac{3}{2} \dfrac{R_{\beta} \eta^2}{\left(1+ \rho \right)^2}\dfrac{\Omega^2}{l^3} \right]  
\left( \xi - \dfrac{l}{(1+ \rho)} \right),
\end{eqnarray}
\begin{equation}
A_4(\Gamma^j) \equiv \Gamma^5 \left( \dfrac{3 \xi R_{\alpha} \Omega^2}{l} \right) 
+ \Gamma^2 \left[  \xi \left( 1 + \rho \right) - l  \right] \dfrac{\Omega^2 R_{\beta}}{l \left( 1+ \rho \right)^2},
\end{equation}
\begin{equation}
A_3(\Gamma^j)=0,
\end{equation}
\begin{eqnarray}
A_2(\Gamma^j) & \equiv & - \Gamma^5 \left\{ 2c^2 R_{\alpha} \left( 1+ \rho \right)^2 
+ \Omega^2 \left[ 7 l R_{\beta} \xi \left(1 + \rho \right) + 3 R_{\alpha} \left( \eta^2 
+ \xi^2 \right) \left(1+ \rho \right)^2 + l^2 R_{\beta} \left(\rho-2 \right) \right] \right\} 
\nonumber \\
& \times & \dfrac{\left[ \xi + \rho \left( l+ \xi \right) \right] }  {2 l^3 \left(1+ \rho \right)^3} 
+ \Gamma^4 \left[ 2 l^2 R_{\beta} (1+ \rho) \right]   \dfrac{\Omega^2 \left[ \xi 
+ \rho \left( l+ \xi \right) \right] }  {2 l^3 \left(1+ \rho \right)^3},
\end{eqnarray}
\begin{equation}
A_1(\Gamma^j) \equiv \Gamma^5 \left( \xi + l \dfrac{\rho}{(1+\rho)} \right) \left(\dfrac{c R_{\alpha}}{l^2} \right)^2,
\end{equation}
\begin{equation}
A_0(\Gamma^j) \equiv \Gamma^5 \left[ \xi + \left( l + \xi \right) \rho \right] 
\dfrac{3 R_{\alpha}\rho \eta^2 \Omega^2 }{2 l^3 \left(1+\rho \right)^3},
\end{equation}
whereas the coefficients $B_n(\Gamma^j)$ are defined by
\begin{eqnarray}
B_5(\Gamma^j) & \equiv & \Gamma^5 \left[ 2 c^2 + \left( \eta^2 + \xi^2 \right) \Omega^2 \right] 
\dfrac{\Omega^2}{2 c^2} + \Gamma^4 \left( \dfrac{3 R_{\beta}\Omega^2}{l} \right)
\nonumber \\
& + & \Gamma^2 \left\{ - \left( 1 + \rho \right)^2 \left[ 2 c^2 + 3 \Omega^2 \left( \eta^2 
+ \xi^2 \right) \right] + \Omega^2 \left[ 7 l \xi \left( 1 + \rho \right) 
+l^2 \left( 2 \rho - 3 \right) \right] \right\} 
\nonumber \\
& \times & \dfrac{R_{\beta}}{2l^3 \left(1+\rho \right)^2} + \Gamma \left(\dfrac{c R_{\beta}}{l^2} \right)^2
+\dfrac{3 R_{\beta}\eta^2 \Omega^2}{2 l^3 \left(1+ \rho \right)^2},
\end{eqnarray}
\begin{equation}
B_4(\Gamma^j) \equiv \Gamma^5 \left( \dfrac{3 R_{\alpha} \Omega^2}{l} \right) 
+ \Gamma^2 \left[ \dfrac{R_{\beta} \Omega^2}{l (1+ \rho)} \right],
\end{equation}
\begin{equation}
B_3(\Gamma^j)=0,
\end{equation}
\begin{eqnarray}
B_2(\Gamma^j)& \equiv & -\Gamma^5 \left\{\dfrac{c^2 R_{\alpha}}{l^3}+ \Omega^2 \left[ \dfrac{7 R_{\beta}\xi}{2l^2(1+\rho)} 
+ \dfrac{3 R_{\alpha}\left( \eta^2 + \xi^2 \right)}{2 l^3}- \dfrac{R_{\beta}}{l (1+\rho)^2} 
+  \dfrac{3 R_{\beta} \rho}{2 l (1+ \rho)^2} \right] \right\} 
\nonumber \\
& + & \Gamma^4 \left[ \dfrac{\Omega^2 R_{\beta}}{l (1+\rho)} \right],
\end{eqnarray}
\begin{equation}
B_1(\Gamma^j) \equiv \Gamma^5 \left( \dfrac{c R_{\alpha}}{l^2} \right)^2,
\end{equation}
\begin{equation}
B_0(\Gamma^j) \equiv \Gamma^5 \left[ \dfrac{3}{2} \dfrac{R_{\beta} \rho \eta^2 \Omega^2}{l^3 (1+\rho^2)} \right].
\end{equation}
The planetoid coordinates are eventually expressed, from the definitions (\ref{(4.7)}) and (\ref{(4.8)}), in the form
\begin{equation}
\xi={l \over 2}\left[(\gamma^{2}-\Gamma^{2})+{(1-\rho)\over (1+\rho)}\right],
\label{(4.18)}
\end{equation}
\begin{equation}
\eta=\pm l \sqrt{\gamma^{2}-{1 \over 4}(\gamma^{2}-\Gamma^{2}+1)^{2}}.
\label{(4.19)}
\end{equation}
By numerical analysis of Eqs. (\ref{(4.16)})-(\ref{(4.17)}) we have found that, in the Earth-Moon system, 
the only solution where both $\gamma$ and $\Gamma$
are different from zero is given by
\begin{equation}
\gamma=0.99999999999996386756, \;
\Gamma=0.99999999999284192083.
\label{(4.20)}
\end{equation}
These values lead to a tiny departure from the equilateral triangle picture of Newtonian 
theory (this effect was first predicted in Ref. \cite{Krefetz1967}), but less
pronounced than in our earlier work \cite{BEDS15}, 
where we found a correction of $8.7894 \; {\rm mm}$ on the $\xi$-coordinate and of 
$-4 \; {\rm mm}$ on the $\eta$-coordinate \cite{BEDS15}. We now find instead, for the planar coordinates of $L_{4}$,
\begin{equation}
\xi^{(GR)}-\xi^{(N)}=2.73 \; {\rm mm}, \;
\eta^{(GR)}-\eta^{(N)}=-1.59 \; {\rm mm}.
\label{(4.21)}
\end{equation}
At this stage, we can compare these corrections with those obtained through the method outlined by the authors of 
Ref. \cite{Yamada}, where the position of the Lagrangian points is obtained by employing the Einstein-Infeld-Hoffman 
equation of motion rather than the analysis of the zeros of the gradient 
of the effective potential $w$ in (4.9). 
As we can see, the correction on the $\xi$-coordinate has got the same sign and the 
same magnitude as the one obtained with the method of Ref. \cite{Yamada}, while the correction on the $\eta$-coordinate 
has got only the same sign, because the magnitude is three times bigger. Therefore, it is interesting to note the fact 
that two different methods give exactly the same correction of the $\xi$-coordinate. By taking account of Eq. 
(\ref{(4.21)}), the resulting values of planetoid distance from Earth and Moon turn out to be
\begin{equation}
r=\gamma l=3.8439999999998611069 \times 10^{8} {\rm m}, \;
s=\Gamma l=3.8439999999724843437 \times 10^{8} {\rm m}.
\label{(4.22)}
\end{equation}

\subsection{Collinear Lagrangian points}

The position of the collinear Lagrangian points $L_1$, $L_2$ and $L_3$ is described by the system of equations
\begin{equation}
\left \{
\begin{array}{ll}
& \dfrac{\partial W}{\partial \xi}=0 \\
& \eta=0,
\end{array} 
\right.
\end{equation}
Following Ref. \cite{BEDS15}, we know that the vanishing of the $\eta$-coordinate implies that
\begin{equation}
\xi = \epsilon r - l \dfrac{\rho}{(1+\rho)}, \; \; \; \; (\epsilon= \pm 1), \label{coll1}
\end{equation}
which in turn leads to the condition 
\begin{equation}
s= \pm (r- \epsilon l). \label{coll2}
\end{equation}
If we substitute relations (\ref{coll1})--(\ref{coll2}) into Eq. (\ref{(4.12)}) and initially adopt the choice 
$s=(r- \epsilon l)$, we obtain an algebraic tenth degree equation where the only unknown is the distance $r$ of 
the planetoid from the Earth. By setting, as before, $r=\gamma l$, this equation can be written down as
\begin{equation}
\sum_{n=0}^{10} C_n \gamma^n = 0, \label{coll3}
\end{equation} 
where
\begin{equation}
C_{10} \equiv 1,
\end{equation}
\begin{equation}
C_{9} \equiv - \dfrac{(7 \rho +4)}{\epsilon (1+ \rho)},
\end{equation}
\begin{equation}
C_{8} \equiv \dfrac{2 c^2}{\Omega^2 l^2} + \dfrac{3 (7 \rho^2 + 8 \rho +2)}{(1+\rho)^2},
\end{equation}
\begin{eqnarray}
C_{7} & \equiv & - \dfrac{1}{\epsilon (1+ \rho)}  \bigg \{ \dfrac{c^2}{l^3 \Omega^2} \left[ 2l (5 \rho +4)
-3 \epsilon (1+ \rho) (R_{\alpha}+R_{\beta}) \right] 
\nonumber \\
&+& \dfrac{1}{(1+ \rho)^2} \left[ \rho^2 (13 \rho + 12)+ 2 (1 + \rho)^2 (11 \rho +2) \right] \bigg  \},
\end{eqnarray}
\begin{eqnarray}
C_{6} & \equiv & \dfrac{c^2}{\epsilon \Omega^2 l^3} \left \{ 12 (l \epsilon -R_{\alpha} - R_{\beta})
+ \rho \left[ 20 l \epsilon - 12  (R_{\alpha}+R_{\beta}) \right] \right \}  
\nonumber \\
&+& \dfrac{1}{(1+ \rho)^3} \left[ 4 \rho^3 + (1 + \rho) (31 \rho^2 + 14 \rho + 1) \right],
\end{eqnarray}
\begin{eqnarray}
C_{5} & \equiv & -2\dfrac{c^4  (R_{\alpha}+R_{\beta}) }{l^5 \Omega^4} + \dfrac{c^2}{l^3 \Omega^2 (1+ \rho)^2} 
\bigg [ - 4 l \epsilon (1 + \rho) (5 \rho +2) + 3 R_{\alpha} (5 \rho^2 +12 \rho +6) 
\nonumber \\
& + & R_{\beta} (18 \rho^2 + 44 \rho + 23) \bigg ] -\dfrac{3 \epsilon \rho}{(1+ \rho)^3} (7 \rho^2 + 6 \rho +1), 
\end{eqnarray}
\begin{eqnarray}
C_{4} & \equiv & \dfrac{2 c^2}{\epsilon l^3 \Omega^2} \bigg \{ [ 2l (R_{\beta}-2R_{\alpha}) 
- \epsilon ((R_{\alpha})^{2}+(R_{\beta})^{2}) ] 
\left( \dfrac{-c^2}{l^3 \Omega^2} \right) + \dfrac{R_{\beta}}{(1+\rho)^2} 
[ 5 \rho^2 + 2 \rho + 5 -2 \epsilon (1 + \rho ) ]   
\nonumber \\
& + & \dfrac{1}{(1+\rho)^2} [ l \epsilon (1+\rho)(1+5 \rho) - 6 R_{\alpha} (1+2 \rho) ] \bigg \} 
+ \dfrac{\rho^2}{(1+\rho)^3} (7 \rho + 3),
\end{eqnarray}
\begin{eqnarray}
C_{3} & \equiv & -\dfrac{2 c^4}{l^6 (1+\rho)^3 \Omega^4 \epsilon} [ R_{\beta}(R_{\beta}+l \epsilon) 
+ R_{\alpha}(4 R_{\alpha}+ 6 l \epsilon)] -\dfrac{c^2}{l^3 (1+ \rho)^2 \Omega^2 \epsilon} \{ 2 l \rho (1+ \rho) 
\nonumber \\ 
&+&  3 R_{\alpha} \epsilon (5 \rho^2 -2 \rho -1) + R_{\beta} [10(1+ \rho) -  \epsilon (3 \rho^2 
+ 44 \rho + 18)] \} -\dfrac{\rho^3}{(1+\rho)^3 \epsilon},
\end{eqnarray}
\begin{equation}
C_{2} \equiv \dfrac{c^2}{l^3 \Omega^2} \left \{ \dfrac{4 c^2 R_{\alpha}}{l^3 \Omega^2} (3 L_{\alpha} 
+ 2 l \epsilon) + \dfrac{1}{(1+\rho)^2} \{ 12 \epsilon R_{\alpha}\rho^2 
-8 R_{\beta} [ \epsilon (1+ 3 \rho) -(1+\rho)] \} \right \}, 
\end{equation}
\begin{equation}
C_{1} \equiv -\dfrac{c^2 \epsilon}{l^3\Omega^2 (1+\rho)^2} \left \{ [2 c^2 R_{\alpha} (4 R_{\alpha}+l \epsilon)(1+\rho)^2] 
\dfrac{\epsilon}{l^3 \Omega^2}+3 \rho^2 R_{\alpha} + 2 R_{\beta} [\epsilon (1+\rho)-(1+3\rho)] \right \},
\end{equation}
\begin{equation}
C_{0} \equiv 2\left(\dfrac{ c^2 R_{\alpha}}{l^3 \Omega^2}\right)^2,
\end{equation}
whereas in the other case, i.e. $s=-(r-\epsilon l)$, we end up with the algebraic equation
\begin{equation}
\sum_{n=0}^{10} D_n \gamma^n = 0, \label{coll4}
\end{equation} 
with
\begin{equation}
D_k=C_k \; \; \;   {\rm if} \; \;  k=10,9,8,0,
\end{equation}
\begin{equation}
D_7 \equiv C_7 - 6 \dfrac{c^2 R_{\beta}}{l^3 \Omega^2},
\end{equation}
\begin{equation}
D_6 \equiv C_6 + 24 \dfrac{c^2 R_{\beta}}{l^3 \Omega^2 \epsilon},
\end{equation}
\begin{equation}
D_5 \equiv C_5 - \dfrac{2 c^2 R_{\beta}}{l^5 \Omega^4 (1+\rho)^2}
[l^2 \Omega^2 (18 \rho^2+38 \rho+21)-2c^2 (1+\rho)^2],
\end{equation}
\begin{equation}
D_4 \equiv C_4 - \dfrac{2 c^2 R_{\beta}}{l^5 \Omega^4 (1+\rho)^2 \epsilon} \{ 4 c^2 (1+\rho)^2 
+ 2 \Omega^2 l^2 [2 \epsilon (1+\rho)-(6 \rho^2+14 \rho+9)] \},
\end{equation}
\begin{equation}
D_3 \equiv C_3 + \dfrac{2 c^2 R_{\beta}}{l^3 \Omega^2} \left [\dfrac{2 c^2 }{l^2 \Omega^2}+\dfrac{10 }
{(1+\rho) \epsilon}-\dfrac{1}{ (1+\rho)^2} (3 \rho^2 + 8 \rho +6) \right], 
\end{equation}
\begin{equation}
D_2 \equiv C_2 -\dfrac{16 c^2 R_{\beta}}{l^3 \Omega^2 (1+\rho)},
\end{equation}
\begin{equation}
D_1 \equiv C_1+\dfrac{4 c^2 R_{\beta}\epsilon}{l^3 \Omega^2 (1+ \rho)}.
\end{equation}
The values $R_i$ ($i=1,2,3$) of the distance of the planetoid from the Earth at the libration points 
$L_1$, $L_2$, $L_3$, respectively, obtained through the solution of Eqs. (\ref{coll3}) and (\ref{coll4}) are given by
\begin{eqnarray}
& R_1= 3.2637628817407598555 \times 10^8 \;  {\rm m}, \label{R1_GR}\\
& R_2=  4.4892056003414800050  \times 10^8 \;  {\rm m},\\
& R_3= 3.8167471569392170594 \times 10^8 \;  {\rm m}, \label{R3_GR}
\end{eqnarray}
whereas the corresponding classical Newtonian values read as \cite{BEDS15} 
\begin{eqnarray}
& r_1=3.263762881738878  \times 10^8 \;  {\rm m},\\
& r_2= 4.489205600344675  \times 10^8 \;  {\rm m}, \\
& r_3=   3.816747156939623  \times 10^8 \;  {\rm m}.
\end{eqnarray}
By comparing these values we have
\begin{equation}
R_{1}^{(GR)}-r_{1}^{(N)}=
R_{1}-r_{1}= 0.19 \; {\rm mm} \; {\rm at} \; L_{1},
\end{equation}
\begin{equation} 
R_{2}^{(GR)}-r_{2}^{(N)}=
R_{2}-r_{2}= -0.32 \; {\rm mm} \; {\rm at} \; L_{2}, 
\end{equation}
\begin{equation}
R_{3}^{(GR)}-r_{3}^{(N)}=
R_{3}-r_{3}= - 0.04 \; {\rm mm} \; {\rm at} \; L_{3}.
\end{equation}
Interestingly, the correction on the position of the Lagrangian point $L_1$ is exactly the same as the one calculated 
with the method described in Ref. \cite{Yamada10}, where\footnote{As shown in Ref. \cite{Yamada10}, the general
relativity corrections to $L_{1},L_{2}$ may be of order $30$ meters in the Sun-Jupiter system. However, 
compared to the Earth-Moon system, a mission to test this effect at Jupiter would be exceedingly more expensive
and complex to realize and could not even benefit from the use of accurate, direct laser ranging from Earth due
to the large distance. The effect of the extremely harsh Jupiter radiation environment on the test spacecraft
(planetoid) should also be considered to evaluate its impact on the integrity of the spacecraft and, therefore,
the duration of the positioning measurements.}
the collinear solutions of the three-body problem are 
studied in the post-Newtonian regime. We believe that, according to the definitions involving the ratio 
of the distances of the planetoid from the primaries given in Ref. \cite{Yamada10}, the equations resulting from the 
application of the method developed by the authors of Ref. \cite{Yamada10} (which is the same method used in Ref. 
\cite{Yamada}) are well suited to describe only the position of $L_1$, and the agreement with the corrections 
presented here is a clue supporting our opinion. 

\section{Quantum effects on Lagrangian points}

The analysis of the previous section prepares the ground for a more appropriate definition and
evaluation of quantum corrections of Lagrangian points, when the underlying classical theory of 
gravity is Einstein's general relativity. For this purpose, we begin by considering the analysis
in Ref. \cite{Brumberg1972}, where the metric tensor components in a co-rotating frame for the relativistic
restricted planar three-body problem in the post-Newtonian limit were obtained. With the notation of our Sec. IV, 
and coordinates $x^{0}=ct,x^{1}=\xi,x^{2}=\eta,x^{3}=\zeta$, the result
in Ref. \cite{Brumberg1972} reads as (cf. Ref. \cite{Krefetz1967})
\begin{eqnarray}
g_{00}&=& 1-2{R_{\alpha}\over r}-2{R_{\beta}\over s}-{\Omega^{2}\over c^{2}}(\xi^{2}+\eta^{2})
+2\left[\left({R_{\alpha}\over r}\right)^{2}+\left({R_{\beta}\over s}\right)^{2}\right]
\nonumber \\
&-& 2 {(R_{\alpha}+R_{\beta})\over l^{3}}
\left({R_{\alpha}\over r}+{R_{\beta}\over s}\right)(\xi^{2}+\eta^{2})
+4{R_{\alpha}\over r}{R_{\beta}\over s} 
\nonumber \\
&+& {(2-\rho)\over (1+\rho)}{R_{\alpha}\over r}{R_{\beta}\over l}
+{(2\rho-1)\over (1+\rho)}{R_{\beta}\over s}{R_{\alpha}\over l}
-7{\xi \over l^{2}}\left({R_{\alpha}\over r}R_{\beta}-{R_{\beta}\over s}R_{\alpha}\right)
\nonumber \\
&+& (1+\rho)^{-1}{\eta^{2}\over l}\left[\rho \left({R_{\alpha}\over r}\right)^{3}
{R_{\beta}\over (R_{\alpha})^{2}}
+\left({R_{\beta}\over s}\right)^{3}{R_{\alpha}\over (R_{\beta})^{2}}\right],
\label{(5.1)}
\end{eqnarray}
\begin{equation}
2c g_{01}=\left(1+2{R_{\alpha}\over r}+2{R_{\beta}\over s}\right)2 \Omega \eta,
\label{(5.2)}
\end{equation}
\begin{equation}
2c g_{02}=-\left(1+2{R_{\alpha}\over r}+2{R_{\beta}\over s}\right)2 \Omega \xi
-8{\Omega^{2}l \over (1+\rho)}\left(\rho {R_{\alpha}\over r}-{R_{\beta}\over s}\right),
\label{(5.3)}
\end{equation}
\begin{equation}
g_{03}=0,
\label{(5.4)}
\end{equation}
\begin{equation}
g_{ij}=-\left(1+2{R_{\alpha}\over r}+2{R_{\beta}\over s}\right)\delta_{ij}, \; i,j=1,2,3.
\label{(5.5)}
\end{equation}
The resulting Lagrangian that describes the planetoid motion in the gravitational field
of Earth and Moon reads as \cite{Wanex,Foster}
\begin{equation}
L={1 \over 2} \sum_{\mu,\nu=0}^{3}
g_{\mu \nu}{{\rm d}x^{\mu}\over {\rm d}t}{{\rm d}x^{\nu}\over {\rm d}t}.
\label{(5.6)}
\end{equation}
We now bear in mind that, in light of second line of (4.1), the dimensionless ratio
\begin{equation}
U_{\alpha}(r) \equiv {R_{\alpha}\over r}=U_{\alpha},
\label{(5.7)}
\end{equation}
where $R_{\alpha} \equiv {G \alpha \over c^{2}}$ is the gravitational radius of the primary
of mass $\alpha$, gets replaced by (or mapped into)
\begin{eqnarray}
{V}_{\alpha}(r) & \sim & 
\left[1+\kappa_{2}{(l_{P})^{2}\over r^{2}}\right]U_{\alpha}(r)
+\kappa_{1}\left(1+{R_{m}\over R_{\alpha}}\right)(U_{\alpha}(r))^{2}
+{\rm O}(G^{3})
\nonumber \\
& \sim & \left[1+\kappa_{2}{(l_{P})^{2}\over r^{2}}\right]U_{\alpha}(r)
+\kappa_{1}(U_{\alpha}(r))^{2},
\label{(5.8)}
\end{eqnarray}
because the gravitational radius $R_{m}$ of the planetoid or laser ranging test mass
is indeed much smaller than $R_{\alpha}$.
The same holds for the dimensionless ratio
\begin{equation}
U_{\beta}(s) \equiv {R_{\beta}\over s}=U_{\beta}
\label{(5.9)}
\end{equation}
and its effective-gravity counterpart
\begin{eqnarray}
{V}_{\beta}(s) & \sim & \left[1+\kappa_{2}{(l_{P})^{2}\over s^{2}}\right]U_{\beta}(s)
+\kappa_{1}\left(1+{R_{m}\over R_{\beta}}\right)
(U_{\beta}(s))^{2}+{\rm O}(G^{3}),
\nonumber \\
& \sim & \left[1+\kappa_{2}{(l_{P})^{2}\over s^{2}}\right]U_{\beta}(s)
+\kappa_{1}(U_{\beta}(s))^{2}.
\label{(5.10)}
\end{eqnarray}
By virtue of Eqs. (5.1)-(5.10), we are led to consider the effective-gravity Lagrangian
\begin{eqnarray}
L_{V}&=& {c^{2}\over 2} \biggr \{ 1-2({V}_{\alpha}+{V}_{\beta})
-{\Omega^{2}\over c^{2}}(\xi^{2}+\eta^{2})
+2 \left[({V}_{\alpha})^{2}+({V}_{\beta})^{2}\right]
\nonumber \\
&-& 2{(R_{\alpha}+R_{\beta})\over l^{3}}(\xi^{2}+\eta^{2})
({V}_{\alpha}+{V}_{\beta})
+4{V}_{\alpha}{V}_{\beta}
\nonumber \\
&+& {(2-\rho)\over (1+\rho)}{R_{\beta}\over l}{V}_{\alpha}
+{(2 \rho-1)\over (1+\rho)}{R_{\alpha}\over l}{V}_{\beta}
-7{\xi \over l^{2}}(R_{\beta}{V}_{\alpha}
-R_{\alpha}{V}_{\beta})
\nonumber \\
&+& (1+\rho)^{-1}{\eta^{2}\over l} \left[\rho {R_{\beta}\over (R_{\alpha})^{2}}
({V}_{\alpha})^{3}
+{R_{\alpha}\over (R_{\beta})^{2}}({V}_{\beta})^{3}\right] \biggr \}
\nonumber \\
&-& {1 \over 2}\Bigr({\dot \xi}^{2}+{\dot \eta}^{2}+{\dot \zeta}^{2}\Bigr)
\Bigr[1+2({V}_{\alpha}+{V}_{\beta})\Bigr]
+\Omega \eta {\dot \xi}\Bigr[1+2({V}_{\alpha}+{V}_{\beta})\Bigr]
\nonumber \\
&-& \Omega \xi {\dot \eta} \Bigr[1+2({V}_{\alpha}+{V}_{\beta})\Bigr]
-4{\Omega^{2}l \over (1+\rho)}{\dot \eta}(\rho {V}_{\alpha}-{V}_{\beta}),
\label{(5.11)}
\end{eqnarray}
and the only nontrivial Euler-Lagrange equations for the planar restricted three-body problem are
\begin{equation}
{{\rm d}\over {\rm d}t}\left({\partial L_{V}\over \partial {\dot \xi}}\right)
-{\partial L_{V}\over \partial \xi}=0, \;
{{\rm d}\over {\rm d}t}\left({\partial L_{V}\over \partial {\dot \eta}}\right)
-{\partial L_{V}\over \partial \eta}=0.
\label{(5.12)}
\end{equation}
Note that, in Refs. \cite{BE14a,BE14b,BEDS15}, we have inserted the effective-gravity map
(see (5.8) and (5.10))
$$
(U_{\alpha},U_{\beta}) \rightarrow ({V}_{\alpha},{V}_{\beta})
$$
in the Lagrangian of Newtonian gravity for the restricted planar three-body problem,
whereas we are here inserting the same map in the Lagrangian of general relativity for
the restricted three-body problem. The metric tensor with components (5.1)-(5.5) describes,
within the framework of general relativity, a tiny departure from the Newtonian treatment of
the restricted planar three-body problem. At that stage, one can recognize that many
Newtonian-potential terms occur therein; for each of them, we apply the effective-gravity
map (5.8) and (5.10) to find what we call a quantum-corrected Lagrangian. 

Note however that in Ref. \cite{BDH2003}, where the authors
derive quantum corrections to some known exact solutions in general relativity, they find that
these metrics differ from the classical metrics only for an additional term proportional to
$(l_{P})^{2}$. Within such a framework, the running of $G$ at large $r$ has a universal character
independent of masses, and there is no room left for $\kappa_{1}$ in the quantum-corrected Lagrangian.
The two schemes are conceptually different: quantum corrections to known {\it exact} solutions of
general relativity do not necessarily have the same nature as quantum corrections of metrics which represent 
solutions of the {\it linearized}  Einstein equations and which are used in turn to derive equations of motion 
of interacting bodies. The insertion of the map (5.8) and (5.10) 
in the Lagrangian of general relativity for three bodies leads to
other terms quadratic in $U_{\alpha}$ and $U_{\beta}$, which are of the same order of those 
already present, and hence the resulting Euler-Lagrange equations (5.12) will lead to predictions
affected by $\kappa_{1}$. 

To further clarify this crucial issue we point out that, if we insert the map (5.8) and (5.10)
in the Lagrangian of Newtonian gravity for the restricted planar three-body 
problem \cite{BE14a}, we find, with our notation, the effective potential
\begin{equation}
W_{\rm eff}={\omega^{2}\over 2}(\xi^{2}+\eta^{2})+c^{2}\Bigr[(U_{\alpha}+U_{\beta})
+\kappa_{1}((U_{\alpha})^{2}+(U_{\beta})^{2})\Bigr]+{\rm O}(G^{2}),
\label{(5.13)}
\end{equation}
whereas general relativity yields the effective potential (4.9), expressible in the form
\begin{eqnarray}
W_{\rm eff}&=&{\Omega^{2}\over 2}(\xi^{2}+\eta^{2})+c^{2}\Bigr[(U_{\alpha}+U_{\beta})
-{1\over 2}((U_{\alpha})^{2}+(U_{\beta})^{2})\Bigr]+{\rm O}(G^{2})
\nonumber \\
& \sim & {\omega^{2}\over 2}(\xi^{2}+\eta^{2})+c^{2}\Bigr[(U_{\alpha}+U_{\beta})
-{1\over 2}((U_{\alpha})^{2}+(U_{\beta})^{2})\Bigr]+{\rm O}(G^{2}),
\label{(5.14)}
\end{eqnarray}
because $\omega^{2}={c^{2}\over l^{3}}(R_{\alpha}+R_{\beta})={\rm O}(G)$ and, by virtue
of (4.4), $\Omega^{2} \sim \omega^{2}+{\rm O}(G^{2})$.
Hence it is possible to understand why the map (5.8) and (5.10) leads to coordinates
of $L_{4}$ and $L_{5}$ in Refs. \cite{BE14a,BE14b,BEDS15} pretty close to those of our Sec. IV.

Moreover, since the work in Refs. \cite{D03,BDH2003} has studied three kinds
of corrected Newtonian potential, i.e. scattering or bound states or one-particle reducible, one
has to rewrite the map (5.8) and (5.10) in the form
\begin{equation}
{V}_{\alpha} \sim U_{\alpha}+\left(\kappa^{\prime}_{1}+{7 \over 2}\kappa_{0}\right)(U_{\alpha})^{2}
+{\rm O}(G^{2}),
\label{(5.15)}
\end{equation}
\begin{equation}
{V}_{\beta} \sim U_{\beta}+\left(\kappa^{\prime}_{1}+{7 \over 2}\kappa_{0}\right)(U_{\beta})^{2}
+{\rm O}(G^{2}),
\label{(5.16)}
\end{equation}
which, upon defining $\kappa_1 \equiv \kappa_1^{\prime} +\dfrac{7}{2} \kappa_0$, takes the form 
\begin{equation}
{V}_{\alpha} \sim U_{\alpha}+ \kappa_1 (U_{\alpha})^{2}
+{\rm O}(G^{2}),
\label{(5.17)}
\end{equation}
\begin{equation}
{V}_{\beta} \sim U_{\beta}+\kappa_1 (U_{\beta})^{2}
+{\rm O}(G^{2}),
\label{(5.18)}
\end{equation}
where
$$
\kappa_{2}{(l_{P})^{2}\over r^{2}}U_{\alpha}={\rm O}(G^{2}),
\kappa_{2}{(l_{P})^{2}\over s^{2}}U_{\beta}={\rm O}(G^{2}),
$$
and the parameter $\kappa_{0}$ vanishes in the scattering and one-particle reducible
cases \cite{BDH2003}, whereas it equals $-1$ for bound states \cite{D03}. Remarkably, since
$\kappa^{\prime}_{1}=3$, and $\kappa_{0}=-1$ for bound states \cite{D03}, this simple calculation shows
that the insertion of the map (5.15) and (5.16) into the Lagrangian of Newtonian gravity 
for the three-body problem leads to the effective potential
\begin{equation}
W_{\rm eff}={\omega^{2}\over 2}(\xi^{2}+\eta^{2})+c^{2}\Bigr[(U_{\alpha}+U_{\beta})
-{1\over 2}((U_{\alpha})^{2}+(U_{\beta})^{2})\Bigr]+{\rm O}(G^{2}),
\label{(5.19)}
\end{equation}
which has the first three terms in common with the effective potential (5.14) of general relativity. 
As far as we can see, this is evidence in favour of inserting the effective-gravity map
into the Lagrangian, and in favour of considering the values of $\kappa_{1}$ and
$\kappa_{0}$ appropriate for bound states both below and in the analysis of Secs. II and III.
Now we set to zero all time derivatives of $\xi$ and $\eta$ in 
Eqs. (5.12), we define the real numbers $\gamma$ and
$\Gamma$ as in Sec. IV and solve numerically the resulting algebraic system for such numbers. This
method leads to the following values:
\begin{itemize}
\item {\bf noncollinear Lagrangian points}:

The planar coordinates of equilibrium points $L_4$ and $L_5$ read as 
\begin{equation}
\xi_4 =\xi_5= 1.8752814880352634039 \times 10^8 \; {\rm m},
\end{equation}
\begin{equation}
\eta_{4,5} = \pm 3.3290016521227759284 \times 10^8 \; {\rm m},
\end{equation}
which means that the differences with respect to the corresponding values provided by the Einstein theory, 
that plays in this scheme the role of the classical theory of reference, read as (cf. Eq. (4.33))
\begin{equation}
\xi_{4} - \xi_{4}^{(GR)}= -1.46 \; {\rm mm}
\Longrightarrow \xi_{4}-\xi_{4}^{(N)}=1.27 \; {\rm mm}, 
\label{5.22}
\end{equation}
\begin{equation}
\eta_{4} - \eta_{4}^{(GR)}=- 0.86 \; {\rm mm}
\Longrightarrow \eta_{4}-\eta_{4}^{(N)}=-2.45 \; {\rm mm}. 
\label{5.23}
\end{equation}

\item {\bf collinear Lagrangian points} :

For the libration points $L_1$, $L_2$ and $L_3$, respectively, we have found that
\begin{equation}
{R}^{\prime}_1=3.2637628817345938976 \times 10^8 \;  {\rm m},  
\end{equation}
\begin{equation}
{R}^{\prime}_2= 4.4892056003375634274 \times 10^8 \;  {\rm m},
\end{equation}
\begin{equation}
{R}^{\prime}_3= 3.81674715692440418189 \times 10^8 \;  {\rm m}. 
\end{equation}
Thus, bearing in mind Eqs. (\ref{R1_GR})--(\ref{R3_GR}), the differences with 
respect to the values expected from general relativity are
\begin{equation}
{R}^{\prime}_1-R_1= -0.62 \; {\rm mm}, \label{5.25}
\end{equation} 
\begin{equation}
{R}^{\prime}_2-R_2= -0.39 \; {\rm mm},
\end{equation} 
\begin{equation}
{R}^{\prime}_3-R_3= -1.48 \;{\rm mm}. \label{5.27}
\end{equation} 
\end{itemize}

Another important issue concerning both noncollinear and collinear Lagrangian points, consists in the fact that 
we have checked numerically that the corrections (\ref{5.22}), (\ref{5.23}) and (\ref{5.25})--(\ref{5.27}) 
do not change if we set $\kappa_2=0$ in the Euler-Lagrange equations (\ref{(5.12)}), because $\kappa_{2}$
weighs the dimensionless ratios ${(l_{P})^{2}\over r^{2}}$ and ${(l_{P})^{2}\over s^{2}}$, which are
extremely small at large values of $r$ and $s$.

\section{General relativity vs. effective gravity: concluding remarks and open problems}

The first part of this paper contributes to make more realistic the model outlined in Refs. 
\cite{BE14a,BE14b,BEDS15} by considering the gravitational presence of the Sun as a perturbing effect for 
the Earth-Moon system. In fact we have shown that also in the quantum regime the presence of the Sun makes 
the planetoid ultimately escape from the triangular libration points, which therefore can be considered as 
``stable'' equilibrium points only during the length of observations. Unless we consider solar radiation pressure, 
from Eqs. (\ref{4aq})--(\ref{4cq}) we have obtained a plot describing the spacecraft motion about $L_4$ 
(Fig. \ref{fig3q}) which is slightly modified if compared with the corresponding classical one (Fig. \ref{fig3c}). 
If we instead take into account the solar radiation pressure (\ref{solar_radiation}), the differences between 
classical and quantum theory become more evident. The presence of solar radiation pressure in the classical case, 
in fact, makes just the planetoid go away from the Lagrangian points $L_4$ more rapidly (see Fig. \ref{fig8c}), 
but in the quantum case, before escaping away from the libration point $L_4$, the planetoid is characterized by 
a less chaotic and irregular motion, as is clear from Fig. \ref{fig8q}. This feature remains true also if we 
consider several initial velocities for the planetoid (Figs. \ref{fig12c} and \ref{fig12q}). 
In particular, we have shown that the reduction of the envelope of the planetoid motion becomes more evident 
in the quantum case. After that, we have 
calculated the impulse needed for the stability of the spacecraft at $L_4$ both in the classical and in the quantum 
regime. These two values, as witnessed by Eq. (\ref{impulsevalue}), are a little bit different and therefore they 
suggest sending two satellites at $L_4$ and $L_5$, respectively, and checking which is the impulse truly needed for 
stability, in order to find out which is, between the classical and the quantum one, 
the best theory suited to describe these phenomena.

In the second part of the paper, we first perform a comparison between Newtonian gravity and general relativity, 
since of course the latter is the most successful theory describing gravitational 
interactions, at least in the solar system. By 
evaluating the points where the gradient of the potential (\ref{(4.9)}) vanishes, we have solved the algebraic equation 
describing the position of Lagrangian points. The distances of noncollinear Lagrangian points from the primaries are 
given in terms of the solutions of Eqs. (\ref{(4.14)})--(\ref{(4.15)}) (or equivalently (\ref{(4.16)})--(\ref{(4.17)})). 
As is clear from Eq. (\ref{(4.21)}), we have obtained corrections of the planar coordinates of the triangular libration 
points of the order of few millimeters and, in particular, the correction on the $\xi$-coordinate is exactly the same 
as the one obtained through the method outlined by the authors of Ref. \cite{Yamada}. As far as collinear Lagrangian 
points are concerned, we have to focus on Eqs. (\ref{coll3}) and (\ref{coll4}), from 
which we have evaluated corrections of the distances of the planetoid from the Earth of the order of few meters. 
In particular, the correction concerning $L_1$ is exactly the same as the one obtainable with the method of Ref. \cite{Yamada10}. 

In the last part of our paper, we have outlined the features of a quantum theory whose underlying classical theory 
is represented by general relativity and not, as before, by Newtonian gravity \cite{D94,D03,BE14a,BE14b,BEDS15}. 
In other words, we have dealt with a theory involving quantum corrections to Einstein gravity, rather than to 
Newtonian gravity. In fact, by applying the map (\ref{(5.8)}) and (\ref{(5.10)}) to the Lagrangian (\ref{(5.6)})  
that general relativity provides for the restricted three-body problem, 
we have ended up with the quantum corrected Lagrangian (\ref{(5.11)}) which, by the means of 
Euler-Lagrange equations (\ref{(5.12)}) together with the conditions 
$\ddot{\xi}=\ddot{\eta}=\ddot{\zeta}=\dot{\xi}=\dot{\eta}=\dot{\zeta}=\zeta=0$, has led us to the few millimeters 
corrections (\ref{5.22}), (\ref{5.23}) and (\ref{5.25})--(\ref{5.27}), obtained by using the quantum 
coefficients $\kappa_1=-1/2$ and $\kappa_2=\dfrac{41}{10 \pi}$ of the bound-states potential discussed in Ref. 
\cite{D03}. We stress that, within this new scheme, quantum corrections on {\it Newtonian} quantities have 
been obtained through the algebraic sum of quantum corrections to general relativity (obtained by (\ref{(5.12)})) 
and general relativity corrections to Newton's theory (obtained by the solutions of Eqs. (\ref{(4.16)}), (\ref{(4.17)}), 
(\ref{coll3}) and (\ref{coll4})). In other words, in this new approach we are no longer using the method developed in Refs. 
\cite{BE14a,BE14b,BEDS15}, where the map (5.8) and (5.10) was only inserted into the Newtonian Lagrangian
of the restricted three-body problem, for simplicity. 
The possibility of mapping the effective potential of Newtonian gravity into an 
effective potential similar to the one of general relativity (cf. (5.19) and (5.14)) adds evidence in favor of the 
choice of $\kappa_{1}$ and $\kappa_{2}$ appropriate for bound states \cite{D03}. We also 
believe it is important to stress that we have used modern packages for dealing with all the coupled algebraic 
equations presented in this paper, verifying eventually that the putative solution does satisfy the original set.

In conclusion, as far as we can see, the implications of calculations presented here and in our previous work are as follows.

\subsection{Noncollinear Lagrangian points}

(i) Quantum corrections to Newtonian planar coordinates of $L_{4}$ and $L_{5}$ and to the corresponding general relativity 
values are (1.27 mm, -2.45 mm) and (-1.46 mm, 0.86 mm), respectively. If it were possible to obtain an
experimental verification of this prediction, it might 
provide encouraging (but not conclusive) evidence in favor of effective theories of gravity
with values $\kappa_{1}=-1/2$ and $\kappa_{2}=41/(10 \pi)$ of the bound-states potential. 
\vskip 0.3cm
\noindent
(ii) If the departure from the equilateral triangle picture of Newtonian theory is instead close to
(2.73 mm, -1.59 mm) for the planar coordinates of $L_{4}$ and $L_{5}$, this may provide another nontrivial
test of classical general relativity.

\subsection{Collinear Lagrangian points}

\begin{table}[!h]
\caption{Distances $r_i$ from the Earth and planar coordinates $(\xi_i,\eta_i)$ of the planetoid at all Lagrangian 
points $L_{i}$ in the classical Newtonian theory, General Relativity and the quantum regime, 
the latter being obtained through the Lagrangian $L_V$ (\ref{(5.11)}) and
with the coefficients of the bound-states potential $\kappa_{1} =-1/2 $ and 
$\kappa_{2} = 41/(10 \pi)$ taken from Ref. \cite{D03}.}
{\relsize{-1.5}
\begin{tabular}{|c|l|l|l|}
\hline
\; \;  $L_i$ \; \;  &  Newtonian Gravity &  General Relativity & Quantum regime \\
\cline{1-4}
& $r_1=3.263762881738878  \times 10^8 \; {\rm m}$& $r_1=3.263762881740760 \times 10^8 \;  
{\rm m}$ & $r_1= 3.263762881734594 \times 10^8 \; {\rm m} $ \\ 
\cline{2-4}
$L_1$ & $\xi_1 = 3.217044369761366\times 10^8 \; {\rm m} $  & $\xi_1 = 3.217044369763247 \times 10^8 \;  
{\rm m}$   & $\xi_1 = 3.217044369757081 \times 10^8 \;  {\rm m}$  \\
& $\eta_1= 0$ &   $\eta_1=0$ &  $\eta_1=0$ \\
\hline
& $r_2=  4.489205600344675 \times 10^8 \; {\rm m} $ & $r_2=4.489205600341480  \times 10^8 \;  
{\rm m}$ & $r_2= 4.489205600337563 \times 10^8 \; {\rm m}$ \\ 
\cline{2-4}
$L_2$ & $\xi_2 = 4.442487088367163 \times 10^8 \; {\rm m}$  & $\xi_2 = 4.442487088363968 \times 10^8 \;  
{\rm m} $  & $\xi_2 = 4.442487088360051  \times 10^8 \;  {\rm m}$  \\
& $\eta_2= 0$ &   $\eta_2= 0$&  $\eta_2= 0$ \\
\hline
& $r_3= 3.816747156939623 \times 10^8 \; {\rm m}$ & $r_3=3.816747156939217 \times 10^8 \;  
{\rm m}$ & $r_3= 3.816747156924404 \times 10^8 \; {\rm m}$ \\ 
\cline{2-4}
$L_3$ & $\xi_3 = -3.863465668917136\times 10^8 \; {\rm m}$  & $\xi_3 = -3.863465668916729 \times 10^8 \; 
{\rm m} $  & $\xi_3 = -3.863465668901917 \times 10^8 \; {\rm m}$  \\
& $\eta_3= 0$ &   $\eta_3= 0$ &  $\eta_3= 0$ \\
\hline
& $r_4= 3.844000000000000\times 10^{8} {\rm m}$ & $r_4= 3.843999999999861 \times 10^{8} 
{\rm m}$ & $r_4 = 3.843999999985078  \times 10^{8} {\rm m}$ \\ 
\cline{2-4}
$L_4$ & $\xi_4 =1.875281488022488  \times 10^8 \;  {\rm m} $  & $\xi_4 = 1.875281488049864  
\times 10^8 \;  {\rm m}$  & $\xi_4 = 1.875281488035263 \times 10^8 \;  {\rm m} $  \\
& $\eta_4=3.329001652147382\times 10^8 \;  {\rm m} $  &   $\eta_4=3.329001652131416 \times 10^8 \;  
{\rm m} $ &  $\eta_4= 3.329001652122776 \times 10^8 \;  {\rm m} $ \\
\hline
& $r_5= 3.844000000000000\times 10^{8} {\rm m}$ & $r_5= 3.843999999999861 \times 10^{8} 
{\rm m}$ & $r_5= 3.843999999985078 \times 10^{8} {\rm m}$ \\ 
\cline{2-4}
$L_5$ & $\xi_5 = 1.875281488022488 \times 10^8 \;  {\rm m} $  & $\xi_5 =1.875281488049864 \times 10^8 \;  
{\rm m}$  & $\xi_5 =  1.875281488035263 \times 10^8 \;  {\rm m}$  \\
& $\eta_5=-3.329001652147382\times 10^8 \;  {\rm m} $ &   $\eta_5= - 3.329001652131416 \times 10^8 \;  
{\rm m}$ &  $\eta_5= - 3.329001652122776 \times 10^8 \;  {\rm m} $ \\
  \hline 
\end{tabular} 
\label{cl_GR_quantum}
}
\end{table} 

\begin{table}[!h]
\caption{Corrections to the distances of $L_1$, $L_2$ and $L_3$ from the Earth and to the planar coordinates of 
$L_4$ and $L_5$. In the first column we have the general relativity corrections to Newtonian theory as obtained from 
Eqs. (\ref{coll3}), (\ref{coll4}), (\ref{(4.16)}) and (\ref{(4.17)}), respectively. The second column shows quantum 
corrections to general relativity given by the Euler-Lagrange equations involving the Lagrangian $L_V$ (\ref{(5.11)}) 
and the quantum coefficients of the bound-states potential. The last column displays quantum corrections to Newtonian values 
calculated as the algebraic sum of the corresponding quantities in the two previous columns.}
{
\begin{tabular}{|l|l|l|l|}
\hline
$L_i$ & General Relativity-Newton & Quantum-General Relativity & Quantum-Newton \\
\hline
$L_1$ & 0.19 mm & -0.62 mm & -0.43 mm \\
\hline
$L_2$ & -0.32 mm & -0.39 mm & -0.71 mm \\
\hline
$L_3$ & -0.04 mm & -1.48 mm & -1.52 mm \\
\hline
$L_4$ & (2.73 mm, -1.59 mm) & (-1.46 mm, -0.86mm) & (1.27 mm, -2.45 mm) \\
\hline
$L_5$ & (2.73 mm, -1.59 mm) & (-1.46 mm, -0.86mm) & (1.27 mm, -2.45 mm) \\
\hline
\end{tabular} 
\label{tab_corrections}
}
\end{table} 
\vskip 0.3cm
\noindent
(iii) If the differences with respect to the classical Newtonian distances of the planetoid at 
$L_1$, $L_2$ and $L_3$ from the Earth were of order (0.19 mm, -0.32 mm, -0.04 mm), 
we would have to consider this fact as another confirmation of general relativity. 
\vskip 0.3cm
\noindent
(iv) In the case in which the quantum theory is ruled by the Lagrangian (\ref{(5.11)}), the quantum corrections 
to general relativity become (-0.62 mm, -0.39 mm, -1.48 mm) at $L_{1},L_{2},L_{3}$, respectively.

A summary of all quantities involved in this paper along with all corrections discussed above is reported for 
clarity in Tabs. \ref{cl_GR_quantum} and \ref{tab_corrections}. As far as we can see, our
detailed calculations show clearly that the measurement we are proposing in the Earth-Moon system 
represents a new testbed for general relativity and effective field theories of gravity. 

\acknowledgments
G. E. is grateful to Dipartimento di Fisica Ettore Pancini of Federico II University for hospitality 
and support. The work of S. D. has been supported by the INFN funding of the 
MoonLIGHT-2 experiment. The authors are grateful
to V. Brumberg, J. F. Donoghue, M. Efroimsky, S. Kopeikin for correspondence, to V. Tiouchov
for help in translating section 8.4 of Ref. \cite{Brumberg1972}, and to E. Calloni for encouragement.

\begin{appendix}

\section{Evaluation of the gradient of $W$ in general relativity}

By virtue of formulas (4.7) and (4.8), one finds
\begin{equation}
{\partial \over \partial \xi}(r^{-p})=-pr^{-p-2}\left(\xi+{\rho l \over (1+\rho)}\right),
\label{(A1)}
\end{equation}
\begin{equation}
{\partial \over \partial \xi}(s^{-p})=-ps^{-p-2}\left(\xi-{l \over (1+\rho)}\right),
\label{(A2)}
\end{equation}
\begin{equation}
{\partial \over \partial \eta}(r^{-p})=-pr^{-p-2},
\label{(A3)}
\end{equation}
\begin{equation}
{\partial \over \partial \eta}(s^{-p})=-ps^{-p-2},
\label{(A4)}
\end{equation}
that we have computed with $p=1,2,3$ to obtain Eqs. (4.12) and (4.13) for the components of
${\rm grad}(W)$, where the functions $W_{1}...W_{4}$ are defined by 
\begin{eqnarray}
W_{1}(\xi,\eta,r)& \equiv & \xi \Omega^{2}+{\Omega^{4}\xi (\xi^2+ \eta^2)\over 2c^{2}}
+3 \Omega^{2}\xi {R_{\alpha}\over r}
+{7 \over 2}{R_{\beta}l \Omega^{2}\over (1+\rho)}{1 \over r}
\nonumber \\
&+& \left(\xi +{\rho l \over (1+\rho)}\right){1 \over r^{3}}
\biggr \{ c^{2}R_{\alpha} \left({R_{\alpha}\over r}-1 \right)+\Omega^{2}
\biggr[-{3 \over 2}R_{\alpha}(\xi^{2}+\eta^{2})
\nonumber \\
&-& {7 \over 2}{R_{\beta}l \xi \over (1+\rho)}
+{3 \over 2}{\rho \over (1+\rho)^{2}}{R_{\beta}l^{2}\eta^{2}\over r^{2}}
+{(2-\rho)\over 2(1+\rho)^{2}}R_{\beta}l^{2}\biggr ] \biggr \} ,
\label{(A5)}
\end{eqnarray}
\begin{eqnarray}
W_{2}(\xi,\eta,s)& \equiv & 3 \Omega^{2}\xi {R_{\beta}\over s}
-{7 \over 2} {R_{\beta}l \Omega^{2}\over (1+\rho)}{1 \over s}
\nonumber \\
&+& \left(\xi -{l \over (1+\rho)}\right){1 \over s^{3}}
\biggr \{ c^{2}R_{\beta} \left({R_{\beta}\over s}-1 \right)+\Omega^{2}
\biggr[-{3 \over 2}R_{\beta}(\xi^{2}+\eta^{2})
\nonumber \\
&+& {7 \over 2}{R_{\beta}l \xi \over (1+\rho)}
+{3 \over 2}{1 \over (1+\rho)^{2}}{R_{\beta}l^{2}\eta^{2}\over s^{2}}
+{(2\rho -1)\over 2(1+\rho)^{2}}R_{\beta}l^{2}\biggr ] \biggr \} ,
\label{(A6)}
\end{eqnarray}
\begin{eqnarray}
W_{3}(\xi,\eta,r)& \equiv & \Omega^{2}+{\Omega^{4}\over 2c^{2}}(\xi^{2}+\eta^{2})
+{c^{2}R_{\alpha}\over r^{3}}\left({R_{\alpha}\over r}-1 \right)
+3 \Omega^{2}{R_{\alpha}\over r}
\nonumber \\
&-& {3 \over 2}\Omega^{2}(\xi^{2}+\eta^{2}){R_{\alpha}\over r^{3}}
-{7 \over 2}{R_{\beta}l \xi \Omega^{2} \over (1+\rho)}{1 \over r^{3}}
\nonumber \\
&+& {R_{\beta}l^{2}\Omega^{2}\over 2(1+\rho)^{2}}{\rho \over r^{3}}
\left(3{\eta^{2}\over r^{2}}-2 \right)
+{R_{\beta}l^{2}\Omega^{2}\over 2(1+\rho)^{2}}
{(2-\rho)\over r^{3}},
\label{(A7)}
\end{eqnarray}
\begin{eqnarray}
W_{4}(\xi,\eta,s)& \equiv & {c^{2}R_{\beta}\over s^{3}}\left({R_{\beta}\over s}-1 \right)
+3 \Omega^{2}{R_{\beta}\over s}
- {3 \over 2}\Omega^{2}(\xi^{2}+\eta^{2}){R_{\beta}\over s^{3}}
+{7 \over 2}{R_{\beta}l \xi \Omega^{2} \over (1+\rho)}{1 \over s^{3}}
\nonumber \\
&+& {R_{\beta}l^{2}\Omega^{2}\over 2(1+\rho)^{2}}{1 \over s^{3}}
\left(3{\eta^{2}\over s^{2}}-2 \right)
+{R_{\beta}l^{2}\Omega^{2}\over 2(1+\rho)^{2}}
{(2 \rho-1)\over s^{3}}.
\label{(A8)}
\end{eqnarray}

\end{appendix}

\end{document}